\documentclass[aps,pra,tightenlines,twocolumn]{revtex4-2}
\usepackage[T1]{fontenc}
\usepackage[utf8]{inputenc}
\usepackage{amsmath}
\usepackage{amsfonts}
\usepackage{amssymb}
\usepackage{graphicx}
\usepackage[english]{babel}
\usepackage{hyperref}

\usepackage{physics}
\usepackage{bm}
\usepackage{slashed}
\usepackage{xcolor}
\usepackage{booktabs}

\graphicspath{{Figures/}}

\newcommand{\R}{\mathbb{R}}

\DeclareMathOperator*{\argmin}{arg\,min}

\DeclareMathOperator{\dist}{dist}
\newcommand{\T}{\mathsf{T}}
\newcommand{\mcal}{\mathcal M}
\newcommand{\mslash}{\slashed{\mathcal M}}

\begin{document}
\title{Thermally Fluctuating, Semiflexible Sheets in Simple Shear Flow}
% Force line breaks with \\

\author{Kevin S. Silmore}
\author{Michael S. Strano}
\author{James W. Swan}
\email{jswan@mit.edu}
\affiliation{Department of Chemical Engineering, Massachusetts Institute of Technology, Cambridge, MA, 02139, USA}

\date{October 20, 2021}

\begin{abstract}
We perform Brownian dynamics simulations of semiflexible colloidal sheets with hydrodynamic interactions and thermal fluctuations in shear flow. As a function of the ratio of bending rigidity to shear energy (a dimensionless quantity we denote $S$) and the ratio of bending rigidity to thermal energy, we observe a dynamical transition from stochastic flipping to crumpling and continuous tumbling. This dynamical transition is broadened by thermal fluctuations, and the value of $S$ at which it occurs is consistent with the onset of chaotic dynamics found for athermal sheets. The effects of different dynamical conformations on rheological properties such as viscosity and normal stress differences are also quantified. Namely, the viscosity in a dilute dispersion of sheets is found to decrease with increasing shear rate (shear-thinning) up until the dynamical crumpling transition, at which point it increases again (shear-thickening), and non-zero first normal stress differences are found that exhibit a local maximum with respect to temperature at large $S$ (small shear rate). These results shed light on the dynamical behavior of fluctuating 2D materials dispersed in fluids and should greatly inform the design of associated solution processing methods.
\end{abstract}

\maketitle

\section{Introduction}

Increasingly, researchers are turning to solution processing methods for 2D materials \cite{stafford2018, ambrosi2018, coleman2011, nicolosi2013, varrla2015, mallory2015, gibaud2017a, ding2017, tardani2018a, davidson2018a, laskar2018, kamal2020, pezzulla2020, klotz2020, liu2018}, yet fundamental knowledge of the dynamical behavior of 2D materials in flow and how such behavior correlates with macroscopic rheological properties and coupled fluid dynamical responses is still lacking. The behavior of rigid ellipsoidal particles suspended in fluids at low Reynolds number was famously studied by Jeffery \cite{jeffery1922}, Hinch and Leal \cite{hinch1972a,hinch1979,leal1971}, and Batchelor \cite{batchelor1970, batchelor1970a} among others. Flexible particles, though, such as polymers \cite{bird1987} or semiflexible sheets, exhibit far more complex behavior compared to their rigid counterparts. In previous work \cite{silmore2021}, we studied the dynamical behavior of thin, \textit{athermal}, semiflexible sheets in simple shear flow as a function of two dominant variables: the initial orientation of a flat sheet about the flow axis, $\phi$, and the dimensionless ratio (denoted $S$) of bending rigidity, $\kappa$, to shear strength given by
\begin{equation}
S = \frac{\kappa }{\pi \eta \dot \gamma L^3},
\label{eq:S}
\end{equation}
where $\eta$ is the solvent viscosity, $\dot \gamma$ is the shear rate, and $L$ is the characteristic radius of the sheet in a flat state.
We observed elastic buckling for sheets oriented initially near the flow-vorticity plane, as well as transitions from quasi-Jeffery behavior to transient tumbling to chaotic, continuous tumbling as $S$ decreases (or as shear strength relative to bending rigidity increases). While the athermal sheets considered in our previous work may serve as a valuable model for macroscopic sheets dispersed in viscous fluids or stiff, large-aspect-ratio nanomaterials like graphene (for which the effective bending rigidity is significantly larger than the thermal energy, $k_B T$), stochastic thermal fluctuations can substantially alter the dynamics. In this work, we sought to explore the effects of thermal fluctuations on the dynamical behavior of sheets under shear at low Reynolds number.

Since the 1980s, the equilibrium and statistical mechanical properties of sheets (or ``tethered membranes'' as they are often called to distinguish them from fluid membranes that lack fixed connectivity) has been widely studied from theoretical \cite{nelson2004, aronovitz1988, frey1991, kantor1986, kardar1988, nelson1987, paczuski1988, kosmrlj2016}, computational \cite{abraham1989, abraham1990, bowick2001, bowick2017, gompper1997, grest1994, ho1989, plischke1988, troster2013}, and experimental \cite{spector1994, wen1992, li2020} viewpoints. Much of this work has focused on a phase transition from a flat state to a crumpled state as $k_B T / \kappa$ is increased, a transition that does not exist for polymers, the 1D analogue of tethered membranes. Another notable difference between polymers and tethered membranes is the lack of an upper critical dimension beyond which self-avoidance effects become negligible ($d_c = 4$ for polymers) \cite{kardar1988}. In fact, most evidence presented to date points to the lack of a crumpling transition for self-avoiding membranes (despite theoretical predictions) and the presence of a rough but flat state at all temperatures \cite{abraham1989, bowick2001, plischke1988, spector1994}.  Additionally, due to thermal fluctuations, thermalized tethered membranes exhibit a wavevector-dependent renormalized bending rigidity, $\kappa_r(q) \sim \kappa (q/q_\mathrm{th})^{-\eta_\kappa}$, where $q$ is the wavevector magnitude and $q_\mathrm{th}$ is the inverse length scale below which thermal fluctuations affect the bending rigidity and $\eta_\kappa \approx 0.8$ is the anomalous dimension \cite{nelson2004}. It should be noted, then, that throughout this work, most results will be reported in terms of the ``bare'' bending rigidity, $\kappa$.

There are still certain unresolved questions about the equilibrium properties of sheets, and there are even more open questions about the dynamics of colloidal sheets, in part due to the few studies that have focused on such a system. Notable previous works on the subject, though, include the following. 
Xu and Green \cite{xu2014, xu2015} studied the behavior of sheets under shear and biaxial extension and found slight shear-thinning at large shear strengths. Babu and Stark \cite{babu2011} studied the fluctuations of tethered sheets in fluids via stochastic rotation dynamics, confirming predicted scaling laws of Frey and Nelson \cite{frey1991} (i.e., that the intermediate scattering function exhibits sub-diffusive scaling that reflects the self-similar roughness of the fluctuating surface). Additionally, Dutta and Graham \cite{dutta2017} classified the dynamical states of athermal Miura-patterned sheets, and Yu and Graham \cite{yu2021} studied ``compact-stretch'' transitions of athermal elastic sheets under extensional flow via the method of regularized Stokeslets.

In this work, we study the behavior of a thermalized ``bead-spring'' sheet model immersed in a low-Reynolds-number simple shear flow. Such sheets can be considered asymptotically thin from the hydrodynamic point of view and are relatively inextensible compared to out-plane bending modes (i.e., they have large Föppl-von Kármán numbers). We conduct Brownian dynamics simulations with hydrodynamic interactions, quantify geometric properties of the sheets, and estimate the viscosity contributions and first normal stress differences as a function of $S$ and the dimensionless temperature, $k_B T / \kappa$. In particular, we look at the orientational covariance matrix of sheet normals, calculate minimum-volume bounding ellipsoids over time, and use such ellipsoids to estimate the aforementioned rheological properties. We find that as $S$ decreases, there is a dynamical transition from intermittent stochastic flipping to continuous tumbling in a crumpled state, in line with our previous study of athermal sheets \cite{silmore2021}.
Finally, scaling predictions for flipping statistics ($\dot \gamma \Delta t_\mathrm{flip} \sim (k_B T / \kappa)^{-1/3} S^{-1/3}$ and $\mathrm{Var}[\dot \gamma \Delta t_\mathrm{flip}] \sim \left( k_B T / \kappa \right)^{-2/3} S^{-2/3}$) are made with the aid of a first passage time model, and all are found to match simulation data well.
%Finally, scaling predictions for the viscosity contribution ($\eta_r \sim (k_B T / \kappa )^{1/3} S^{1/4}$) and flipping statistics ($\dot \gamma \Delta t_\mathrm{flip} \sim (k_B T / \kappa)^{-1/3} S^{-1/3}$ and $\mathrm{Var}[\dot \gamma \Delta t_\mathrm{flip}] \sim \left( k_B T / \kappa \right)^{-2/3} S^{-2/3}$) are made, the latter with the aid of a first passage time model, and all are found to match simulation data well.

\section{Model and Methods}

Similar to the model employed in our previous work \cite{silmore2021}, hexagonal sheets of circumradius $38a$ were constructed by creating a surface triangulation with edges of length $l=2a$ for a total of $N=1141$ vertices or ``beads''. The length scale $a$, here, is the smallest coarse-grained length scale at which relative motion and thermal fluctuations of the sheet are resolved.

Here, we summarize briefly the forces acting between the beads (see \cite{silmore2021} for further details). Bending forces were captured by dihedral forces over each pair of neighboring triangles $\bigtriangleup_i$ and $\bigtriangleup_j$ of the sheet surface as:
\begin{equation}
U_\mathrm{bend}(\bigtriangleup_i, \bigtriangleup_j) = \kappa (1 - \vu n_i \vdot \vu n_j),
\end{equation}
where $\kappa$ is the bending rigidity and $\vu n_i$ and $\vu n_j$ are consistently oriented triangle normals \cite{bowick2017, bian2020, guckenberger2017, gompper1996}. This ``discrete'' value of $\kappa$ can be mapped to a bending rigidity of an equivalent \textit{continuum} sheet, $\tilde \kappa$, via $\tilde \kappa = \kappa / \sqrt{3}$ \cite{gompper1996}.
Harmonic bonds of the form
\begin{equation}
U_\mathrm{bond}(r) = \frac{k}{2} (r - r_0)^2
\end{equation}
were applied between connected beads of the triangulation with stiffness $k = (1000/58) \times 6\pi\eta\dot \gamma L$ and $r_0 = 2a$, where $\eta$ and $\dot \gamma$ are the viscosity and shear rate of the surrounding fluid, respectively. Like the bending rigidity, this value of $k$ can be mapped to a 2D Young's modulus, $Y$, of an equivalent continuum sheet via $Y = 2 k / \sqrt{3}$ \cite{bowick2017}. All of the sheets considered consequently exhibited much softer out-of-plane bending compared to in-plane stretching, with large Föppl-von Kármán (FvK) numbers ($\mathrm{ FvK} = YL^2 / \kappa$) between $10^4$ and $10^6$. Consequently, we do not consider the FvK number to be a relevant dimensionless group for the phenomena observed.

Hard-sphere interactions between all non-neighboring beads were approximated via the pair potential,
\begin{equation}
U_\mathrm{HS}(r) = \begin{cases}
\frac{16 \pi \eta  a^2 \qty[2a\ln(\frac{2a}{r}) + (r - 2a)]}{\Delta t} & 0 \leq r \leq 2a \\
0 & r > 2a
\end{cases},
\end{equation}
where $r$ is the distance between two interacting particles, and $\Delta t$ is the integration timestep used. This pairwise potential displaces two overlapping particles to contact under Rotne-Prager-Yamakawa dynamics (see below) with the same integration timestep. Similar potentials have been employed in modeling hard-sphere fluids \cite{heyes1993} and hard-sphere chain fluids \cite{silmore2020}.

\begin{figure*}[ht]
\centering
\includegraphics{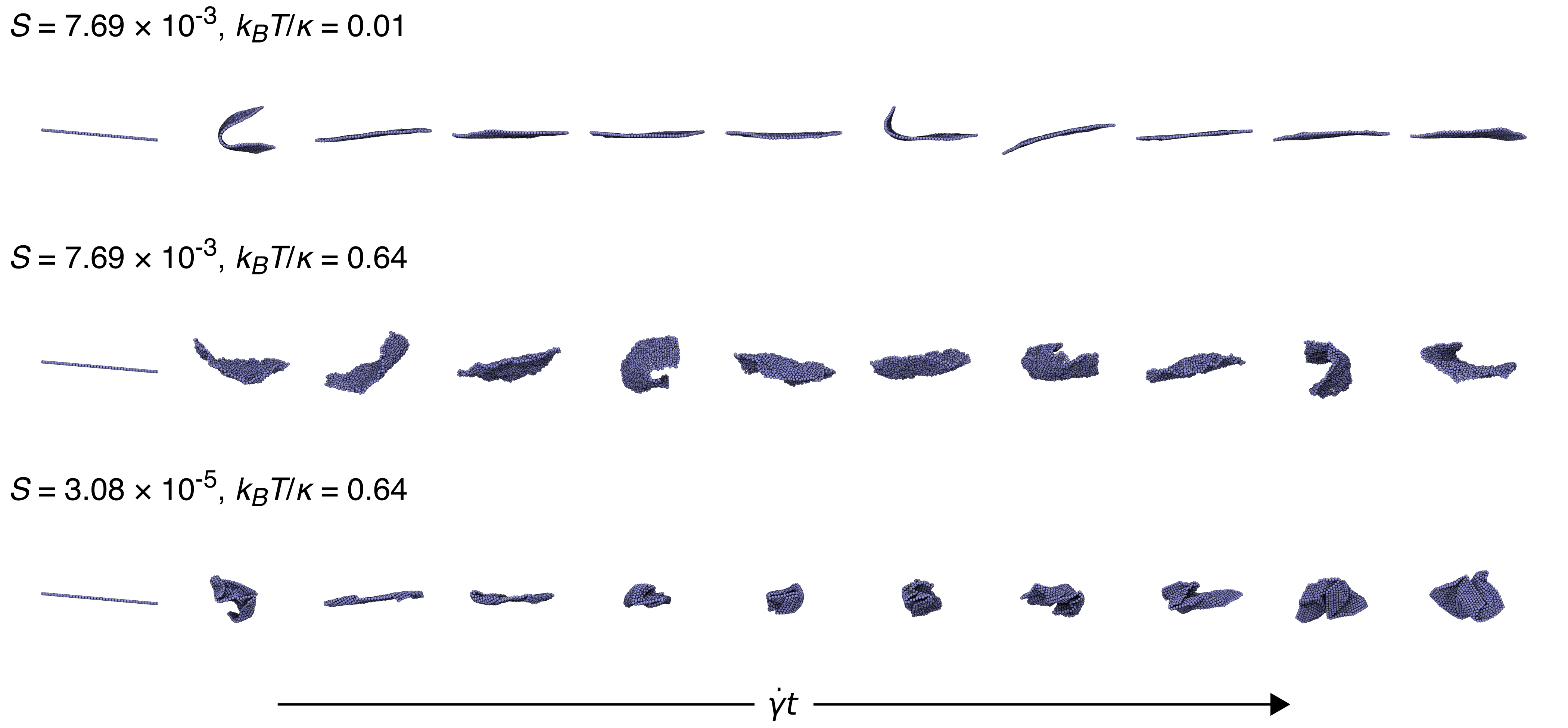}
\caption{Snapshots of sheets every $\dot \gamma t = 10$ units of time for selected dimensionless bending rigidities, $S$, and dimensionless temperatures, $k_B T / \kappa$. The softer sheet (bottom) row crumples unlike the stiffer sheet (top two rows), which only flips stochastically, regardless of temperature.}
\label{fig:snapshots}
\end{figure*}

Brownian dynamics with hydrodynamic interactions was employed to model the motion of all of the beads with the following governing stochastic differential equation:
\begin{equation}
\begin{split}
\dd{ \vb x_i} &= \qty( -\bm{\mcal}_{i j} \partial_j U + \vb L  \vb x_i + k_B T \partial_j \vdot \bm{\mcal}_{ij} )\dd{t} \\
&\quad + \sqrt{2k_B T} \bm{\mslash}_{ij} \dd{\vb W_j},
\end{split}
\label{eq:main_int}
\end{equation}
where $i$ and $j$ are bead indices ranging from 1 to $N$ and map to blocks of size 3, $\vb x \in \R^{3N}$ is a stacked vector of $N$ particle coordinates, ``$\partial_j$'' denotes $\partial/\partial \vb x_j$, $U = U_\mathrm{bend} + U_\mathrm{bond} + U_\mathrm{HS}$ is the total potential energy, and $(\vb L)_{mn} = \dot \gamma\delta_{m1}\delta_{n2}$ is the $3 \times 3$ velocity gradient tensor. Repeated indices are summed over. $\bm{\mcal}$ is the mobility tensor, which maps forces on particle $j$ to particle $i$ and accounts for fluid-mediated interactions between the particles. We set the mobility tensor equal to the well-known Rotne-Prager-Yamakawa (RPY) tensor \cite{rotne1969, yamakawa1970}, which is given analytically by:
\begin{equation}
\bm{\mcal}_{ij} = 
\frac{1}{6 \pi \eta a}
\begin{cases}
\qty(\frac{3a}{4r} + \frac{a^3}{2r^3}) \vb I + \qty(\frac{3a}{4r} - \frac{3a^3}{2r^3})\vu r \vu r^\T & r > 2a \\
\qty(1 - \frac{9r}{32a}) \vb I + \frac{3r}{32a} \vu r \vu r^\T & r \leq 2a
\end{cases},
\end{equation}
where $r$ is the distance between particles $i$ and $j$, and $\vu r$ is a unit vector pointing from particle $i$ to particle $j$. For the RPY tensor, the divergence term $\partial_j \vdot \bm{\mcal}_{ij}$ is equal to 0. Finally, $\bm{\mslash}$ is a matrix that satisfies $\bm{\mslash} \bm{\mslash}^\T = \bm{\mcal}$, and $\vb W$ is a vector of independent, standard Wiener processes. The last term of equation \ref{eq:main_int} represents thermal fluctuations, and the action of $\bm{\mslash}$ is computed efficiently via a Lanczos iteration procedure \cite{fiore2017, chow2014}. The RPY tensor accounts for far-field, pairwise hydrodynamic interactions between two spheres in an unbounded domain and is able to serve as a ``regularized'' Stokeslet for composite-bead objects \cite{fiore2017, swan2016} such as the sheets studied in this work. Hydrodynamically, the sheets in this work behave as if they were asymptotically thin because the beads are constructed to lie on a 2D manifold. Although this leads to Jeffery orbits of infinite period, thermal fluctuations in this work represent the dominant cause of sheet flipping and would remain so even for sheets of finite, but small relative thickness $h / L$. Additionally, the RPY beads that comprise the sheet can be considered an approximation of a ``no-slip'' surface, which means the effects of slip on the dynamics (which can be important for small nanomaterials \cite{kamal2020, kamal2021}) are neglected.

Equation \ref{eq:main_int} was integrated via an Euler-Maruyama scheme with a timestep of $\dot \gamma \Delta t = 5\times10^{-4}$ using a custom plugin adapted from ref. \cite{fiore2017} for the HOOMD-blue molecular simulation package \cite{anderson2020} on graphics processing units (NVIDIA GTX 980s and 1080s). Practically, values of $\kappa$ and $k_B T / \kappa$ were varied while $\dot \gamma$ was set to 1 for the given timestep. The initial conformation for each simulation was a flat sheet rotated by $\theta = 5^\circ$ about the vorticity axis from the flow-vorticity plane, and 6 independent runs of length $\dot \gamma t  =1000$ for each set of parameters were conducted with different random seeds.

\section{Flipping Behavior}

For rigid, athermal, and axisymmetric ellipsoidal particles, the period of the Jeffery orbit is determined by the aspect ratio of the particle \cite{jeffery1922}, and thermal fluctuations cause the particles to diffuse across different orbits \cite{leal1971}. With flexible sheets as well, thermal fluctuations cause sheets to interact with different streamlines of the shear flow and stochastically flip. Figure \ref{fig:snapshots} shows snapshots of sheets with different values of dimensionless bending rigidity, $S$, and different dimensionless temperatures, $k_B T / \kappa$. Additional movies of sheets can be found in the Supporting Information. In general, the stiffest sheets examined ($S \gtrapprox 10^{-3}$) fluctuate about the flat state in the flow-vorticity plane and flip at stochastic intervals while the softest sheets ($S \lessapprox 10^{-4}$) crumple and tumble continuously.

\begin{figure*}[ht]
\centering
\includegraphics{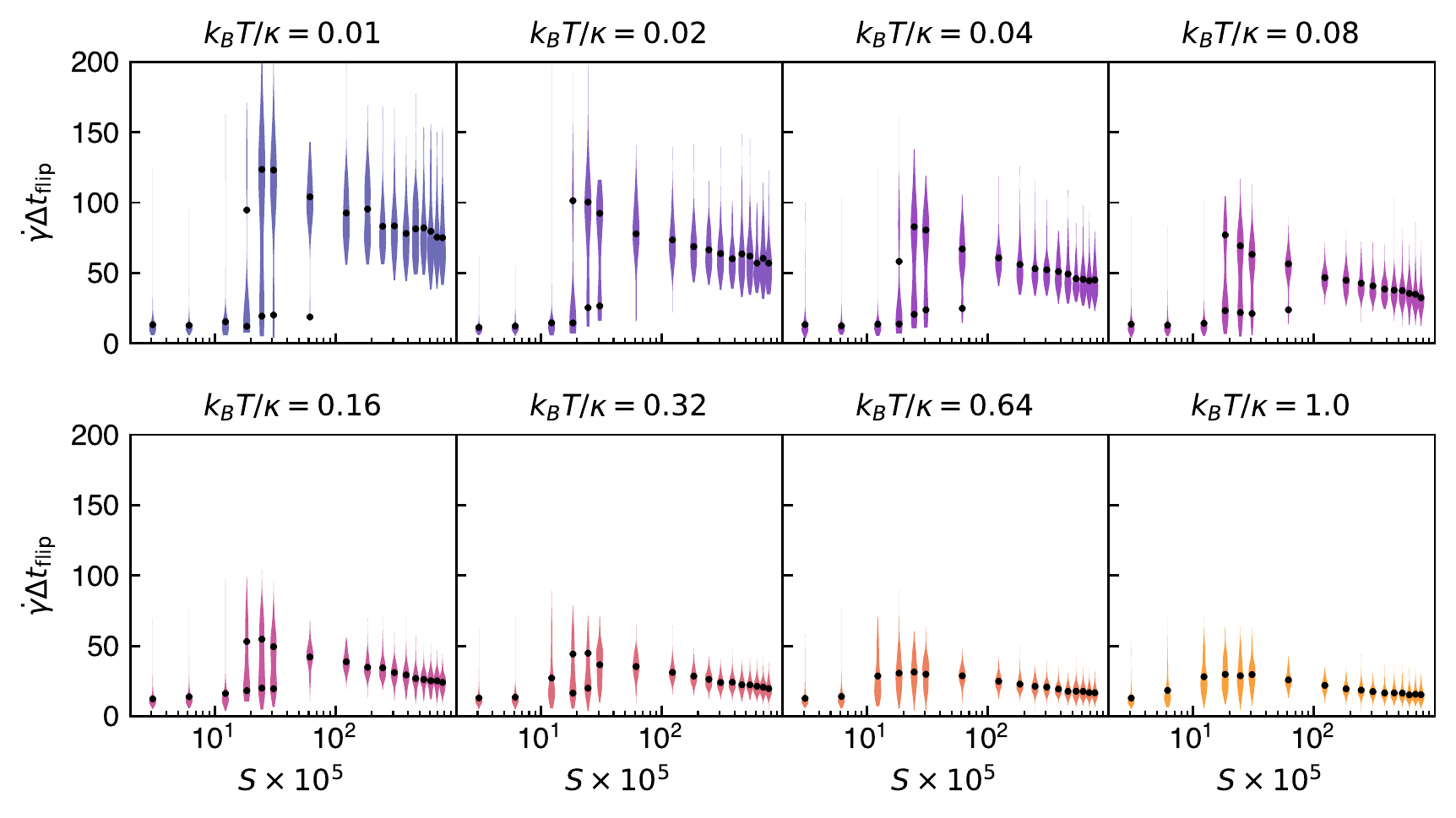}
\caption{The distribution of times between flips (as measured by peak in the bending energy) as a function of dimensionless bending rigidity, $S$, and dimensionless temperature, $k_B T / \kappa$. Dots represent the empirical mean of each distribution or the means of a truncated Gaussian mixture model if its likelihood was relatively larger (see text).}
\label{fig:flipping_times_multi}
\end{figure*}

More quantitatively, Figure \ref{fig:flipping_times_multi} shows the distribution of times between flips, $\dot \gamma \Delta t_\mathrm{flip}$, for the entire range of $S$ values considered in this work and at different temperatures. Specifically, these ``times between flips'' were calculated as the time between distinct peaks in the bending energy. Bending energies (i.e., the sum of all dihedral energies) were calculated every $\dot \gamma \Delta t_\mathrm{log} = 0.1$ units of time, and peaks were automatically extracted from all sheet trajectories by 1) smoothing the data via discrete convolution with a Gaussian of variance $\sigma^2 = 10$ as:
$$\tilde E(\dot \gamma t_i) = \frac{1}{\dot \gamma \Delta t_\mathrm{log} \sqrt{2 \pi \sigma^2}} \sum_j E(\dot \gamma t_j) \exp(\frac{-(\dot \gamma t_i - \dot \gamma t_j)^2}{2 \sigma^2})$$
and 2) identifying peaks with manually tuned thresholds (i.e., that the quadratic coefficient of a quadratic fit with neighboring points within $\pm \dot \gamma \Delta t = 1$ exceeded $0.01$, that the peak height above $\min \tilde E$ was greater than $0.1 (\max \tilde E - \min \tilde E)$, and that the distance in time from the last peak was at least $\dot \gamma t = 2$, which is much smaller than the time to flip from the initial orientation). 

From this data presented in Figure \ref{fig:flipping_times_multi}, one can glean several conclusions. First, the time between flips, in general, decreases as temperature increases. This behavior should not be surprising, since with larger amplitude fluctuations, there is a larger probability that the sheet will interact with a streamline that induces a sufficiently strong drag force to cause the sheet to flip. Second, the flipping time distributions are much wider for smaller dimensionless temperatures, which is consistent with stochastic first-passage-time-like behavior that will be discussed further below. Third, and perhaps most interestingly, there appears to be a \textit{discontinuous} transition to a crumpled state with fast flipping times as $S$ decreases. That these fast flipping times are indeed associated with a geometrically crumpled state will be discussed shortly. In fact, for temperatures less than approximately $k_B T / \kappa = 0.64$, this transition from stochastic flipping to continuous tumbling manifests itself as the appearance and disappearance of discrete modes as $S$ is varied. Without \textit{a priori} knowledge of the distribution of these flipping times, a truncated Gaussian mixture model was fitted to the data for each $S$ with the EM algorithm \cite{bishop2006}. The means of such a model for values of $1.5 \times 10^{-4} < S < 10^{-3}$ are displayed in Figure \ref{fig:flipping_times_multi} if the Bayesian Information Criterion (BIC) difference between a bimodal mixture or single truncated Gaussian was less than -2 as a way of quantifying the discrete modes. For higher temperatures, the transition is much more gradual, and the different flipping and continuously tumbling modes are indistinguishable. The location of the transition around $S \approx 2 \times 10^{-4}$ is consistent with the crumpling/chaos transition found for athermal sheets in our previous work \cite{silmore2021}, and the fact that the transition becomes more ``rounded'' around this athermal crumpling transition is consistent with previous work on the rounding of the buckling transition for flexible filaments \cite{manikantan2015}. It should be noted, though, that the nature of this crumpling/chaos transition is more complex than a buckling transition, the first of which occurs at much larger value of $S \approx 5.3 \times 10^{-3}$ for athermal sheets oriented near the flow-vorticity plane \cite{silmore2021}. This complexity somewhat precludes the typical eigenfunction decomposition and linear stability analysis used to analyze dynamical transitions. Consequently, a more rigorous analytical theory describing such thermal ``rounding'' in future work would be quite valuable.

\begin{figure}[ht]
\centering
\includegraphics[width=\linewidth]{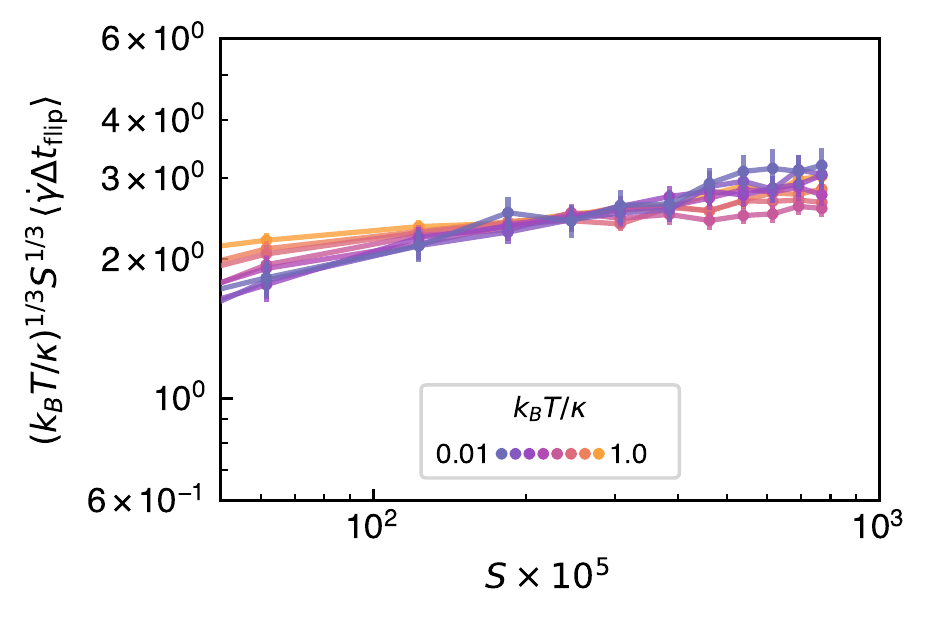}
\caption{Scaled mean time between flips (see Figure \ref{fig:flipping_times_multi}). Error bars represent two standard errors of the means across data from all independent runs, and lines are drawn to guide the eye.}
\label{fig:flipping_times_master}
\end{figure}

Flipping/tumbling frequencies of semiflexible polymers have been studied both experimentally \cite{schroeder2005, smith1999, gerashchenko2006} and theoretically \cite{winkler2006, huang2011, dalal2012}. Notably, it has been found that polymer tumbling in the flow-gradient plane for stiff polymers or large Weissenberg numbers (where $\mathrm{Wi} = \dot \gamma \tau_1$ and $\tau_1$ is the longest relaxation time) is determined by diffusion to a critical angle, $\theta_c$, followed by advection-dominated flipping. Furthermore, $\theta_c \sim \mathrm{Wi}^{-1/3}$ and $D_r \Delta t_\mathrm{diff} \sim \mathrm{Wi}^{-2/3}$, where $D_r$ is the rotational diffusivity of the stiff polymer. A similar analysis can be performed for sheets at large values of $S$ and small values of $k_B T / \kappa$, such that bending rigidity is sufficiently large compared to both the shear strength and thermal energy. Unlike the typical analysis for polymers or dumbbells, motion in all directions --- not just in the flow-gradient plane --- could be especially relevant for a sheet. For small deviations away from the athermal flat state in the flow-vorticity plane, Jeffery's equations for an asymptotically thin, oblate spheroid become:
\begin{equation}
\begin{split}
\dot \theta &= \dot \gamma \theta^2 \\
\dot \phi &= \dot \gamma \phi \theta,
\end{split}
\end{equation}
where $\theta$ is the azimuthal angle between the sheet and the flow-vorticity plane, and $\phi$ is the elevation angle from the flow-gradient plane. $(\theta, \phi) = (0^\circ, 0^\circ)$ corresponds to the athermal flat state of the sheet lying in the flow-vorticity plane. Solving these equations with a (deliberately notated) initial condition of $(\theta_c, \phi_c)$ yields,
\begin{equation}
\begin{split}
\theta(t) = \theta_c (1 - \theta_c \dot \gamma t)^{-1} \\
\phi(t) = \phi_c (1 - \theta_c \dot \gamma t)^{-1},
\end{split}
\end{equation}
indicating an advective time scale of $(\theta_c \dot \gamma)^{-1}$. It is also clear that motion in the $\phi$-direction does not introduce any additional time scales to the problem. Thus, balancing this advective time scale with the characteristic diffusive time scale, $\theta_c^2 / D_r$, yields the familiar scaling:
\begin{equation}
\theta_c \sim (\dot \gamma / D_r)^{-1/3}, \qquad \dot \gamma \Delta t_\mathrm{diff} \sim (\dot \gamma / D_r)^{1/3}.
\label{eq:thetac_scaling}
\end{equation}
Now, given that $D_r \sim k_B T / (\eta L^3)$ for a sheet of characteristic radius $L$, it is possible to rewrite the scaling for $\dot \gamma \Delta t_\mathrm{diff}$ as
\begin{equation}
\dot \gamma \Delta t_\mathrm{diff} \sim \left( k_B T / \kappa \right)^{-1/3} S^{-1/3}.
\label{eq:tdiff_scaling}
\end{equation}
Furthermore, following Harasim et al. \cite{harasim2013}, the time between flips can be represented as the sum of two contributions as $\dot \gamma \Delta t_\mathrm{flip} = \dot \gamma(\Delta t_\mathrm{diff} + \Delta t_\mathrm{adv})$, where $\Delta t_\mathrm{adv}$ is the time required for the sheet to flip due to advective motion from the imposed flow after reaching the critical locus of orientational angles (i.e., all of critical values of $(\theta, \phi)$). For small $\theta_c$, the advective time should scale with $1/\theta_c$ from Jeffery's equations, implying that $\dot \gamma \Delta t_\mathrm{adv}$, and ultimately $\dot \gamma \Delta t_\mathrm{flip}$ as well should scale as $(k_B T / \kappa)^{-1/3} S^{-1/3}$.

Figure \ref{fig:flipping_times_master} shows the (empirical) mean time between flips scaled by $(k_B T / \kappa)^{1/3} S^{1/3}$, which should be constant with $S$ in the realm of applicability of the above analysis. And indeed, for the smallest temperatures and largest values of $S$ plotted in Figure \ref{fig:flipping_times_master}, the data is approximately constant with $S$ and consistent with the proposed scaling. That the scaled mean time between flips deviates from a constant value for both small values of $S$ and larger relative temperatures is unsurprising. The proposed scaling argument is only valid for asymptotically stiff sheets undergoing small thermal fluctuations. More flexible sheets and sheets that adopt more corrugated configurations due to larger thermal fluctuations should flip \textit{more readily} and exhibit \textit{smaller} times between flips given that fluctuations out of plane should enhance interaction with the shear flow.
%In other words, with smaller times between flips, the effective rotational diffusivity of the sheet increases.

\begin{figure}[ht]
\centering
\includegraphics[width=\linewidth]{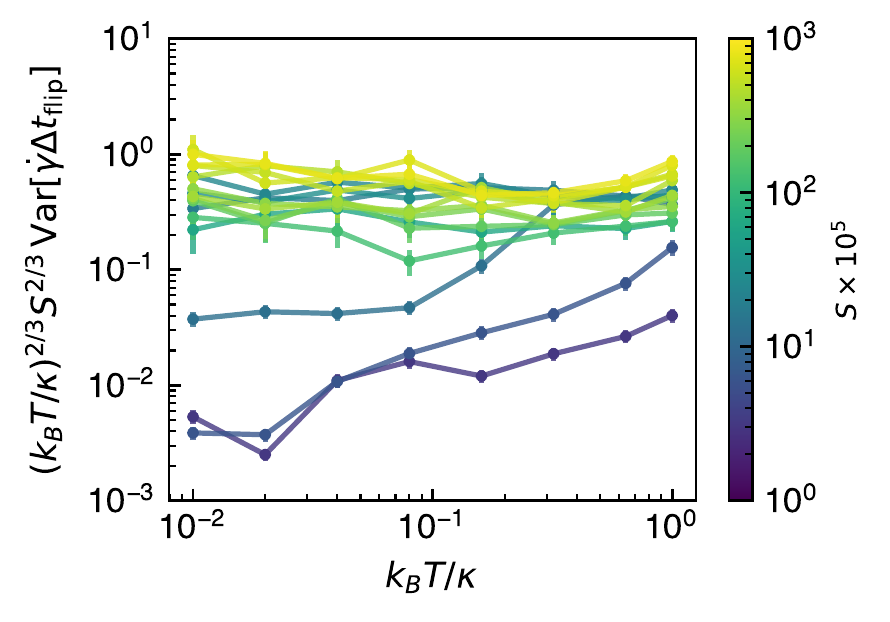}
\caption{Scaled variance of the time between flips (see Figure \ref{fig:flipping_times_multi}). Error bars represent two standard errors across data from all independent runs, and lines are drawn to guide the eye.}
\label{fig:flipping_times_variance}
\end{figure}

\begin{figure*}[ht]
\centering
\includegraphics{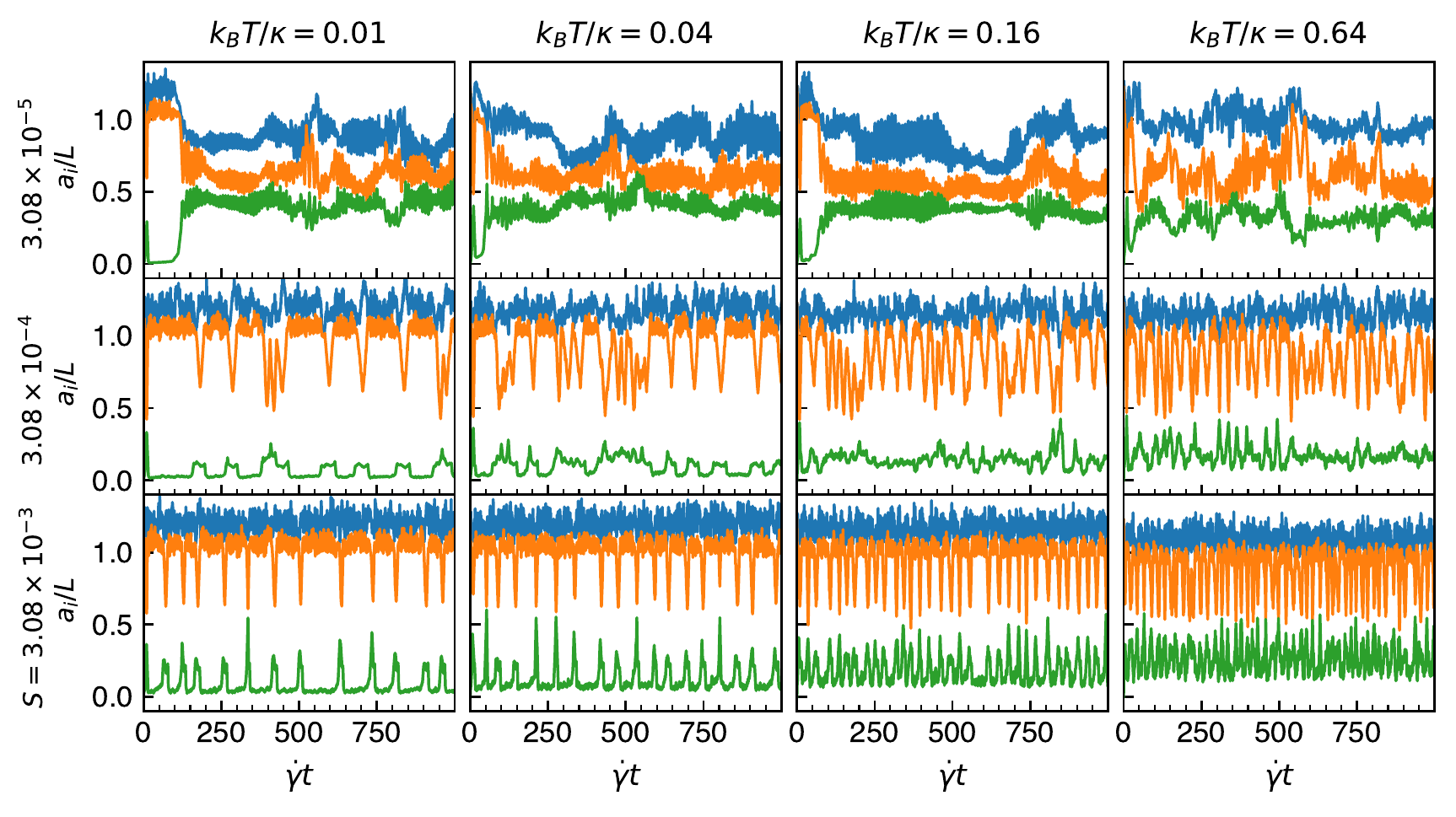}
\caption{The length of the minimum-volume bounding ellipsoid semiaxes (smallest in green and largest in blue) over time during a single run for select values of the dimensionless bending rigidity, $S$, and dimensionless temperature, $k_B T / \kappa$.}
\label{fig:ellipsoid_multi}
\end{figure*}

Considering that rigid sheets aligned in the flow-vorticity plane and subjected to small thermal fluctuations need to diffuse to the locus of critical orientations before advection dominates, it is natural to consider a first passage time model to understand both the mean of the time between flips as well as the variance of the time between flips. Consider a Brownian particle initially located at $x = 0$ and diffusing in the domain $x \in [-\theta_c, \theta_c]$ with diffusivity $D_r$. Let $x = -\theta_c$ be a reflecting (Neumann) boundary condition, and let $x = \theta_c$ represent an absorbing (Dirichlet) boundary condition. Clearly, such a particle is meant to represent the orientation of the sheet, and the boundary conditions model the interaction of the sheet with the flow, with absorption at the right representing the transition from diffusion to advection. As shown in Appendix A, both the mean first passage time to the right boundary and the variance about the mean can be calculated analytically as
\begin{equation}
\mathbb E[\dot \gamma \Delta t_\mathrm{diff}] = \frac{3\dot \gamma \theta_c^2}{2D_r},
\end{equation}
and
\begin{equation}
\mathrm{Var}[\dot \gamma \Delta t_\mathrm{diff}] = \frac{5 \dot \gamma^2 \theta_c^4}{2 D_r^2}.
\end{equation}
Additionally, the flipping time probability distribution function in Appendix A exhibits an exponential tail at long times, which is consistent with the statistical analysis of Chertkov et al. \cite{chertkov2005}. Using equation \ref{eq:thetac_scaling} and assuming the advective time contributes negligibly to the variance, the variance of the time between flips of a rigid sheet rotating to a critical locus at which advection dominates should scale as
\begin{equation}
\begin{split}
\mathrm{Var}[\dot \gamma \Delta t_\mathrm{flip}] &\sim \dot \gamma^2 \frac{\theta_c^4}{D_r^2} \sim \left( \frac{\dot \gamma}{D_r} \right)^{2/3} \\
&\sim \left( k_B T / \kappa \right)^{-2/3} S^{-2/3}.
\end{split}
\end{equation}
Figure \ref{fig:flipping_times_variance} shows the empirical variance of the flipping times illustrated in Figure \ref{fig:flipping_times_multi} scaled by $( k_B T / \kappa)^{2/3} S^{2/3}$, and one can see that data indeed collapses well for the largest values of $S$ for which the theory is appropriate. For the smallest values of $S$ (the softest sheets), one can see that the variance in flipping times is much smaller than predicted by the above first passage time analysis. This is consistent with the fact that those sheets crumple and continuously tumble in a manner that breaks the underlying physical assumptions of the analysis. Overall, though, Figures \ref{fig:flipping_times_master} and \ref{fig:flipping_times_variance} demonstrate that both the model of a critical locus of orientations (separating diffusive first passage and advective motion) and the associated scaling predictions accurately capture the behavior of the stiffest sheets subjected to weak thermal fluctuations.

\section{Conformational Behavior}

\begin{figure*}[ht]
\centering
\includegraphics{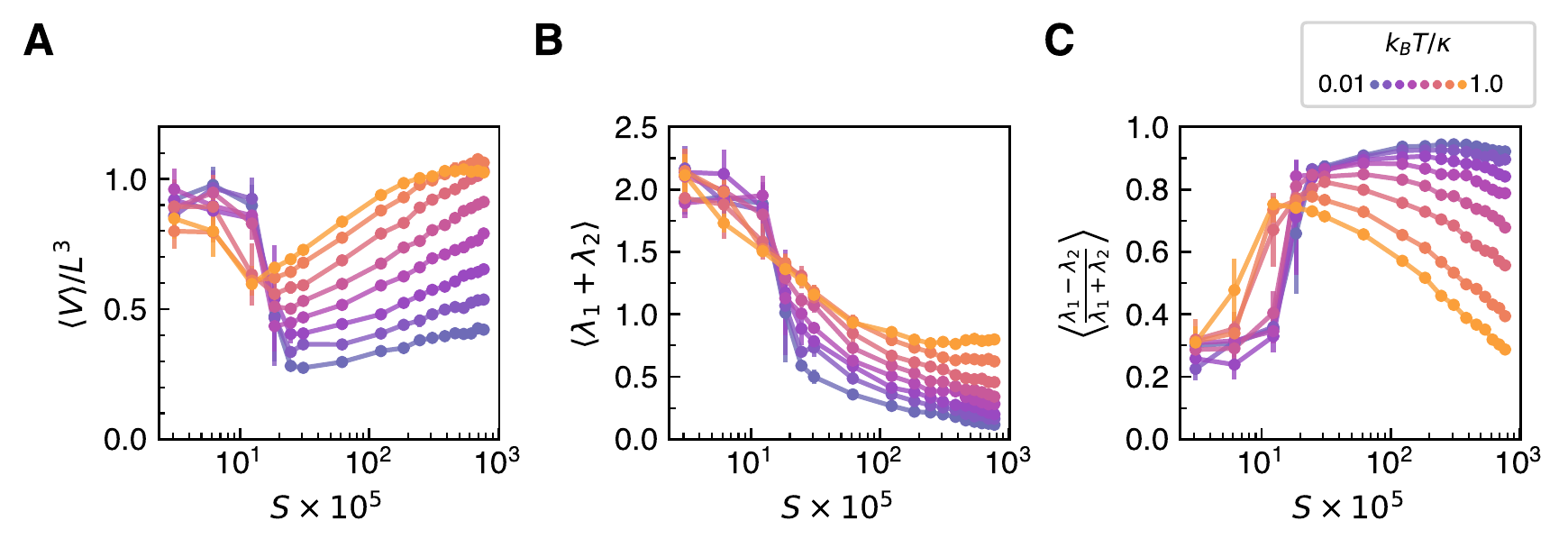}
\caption{\textbf{A)} Time-averaged volume of the minimum-volume bounding ellipsoids. \textbf{ B-C)} The time-averaged sum and normalized difference, respectively, of the eigenvalues of the orientational covariance matrix of the sheets' normals. Error bars in all plots represent two standard errors of the mean between the independent runs, and lines are drawn to guide the eye.}
\label{fig:volume_covariance}
\end{figure*}

As $S$ decreases, there is a transition from stochastic flipping to crumpling/continuous tumbling that is consistent with the transition to chaotic crumpling found for athermal sheets in our previous work \cite{silmore2021}. In order to study the conformational behavior of sheets as a function of $S$ and $(k_B T / \kappa)$, we consider two geometric quantities: minimum-volume bounding ellipsoids and orientational covariance matrices. Minimum-volume bounding ellipsoids of sheets were calculated for snapshots of sheet configurations every $\dot \gamma \Delta t_\mathrm{snap} = 0.25$ units of time via the following (dual) convex optimization problem \cite{kumar2005}:
\begin{equation}
\begin{array}{rl}
\displaystyle \max_{\vb u} & \log \det \left(\sum_{i = 1}^N u_i \vb q_i \vb q_i^\T + \delta \vb I \right)\\
\mathrm{s.t.} & \vb 1^\T \vb u = 1 \\
 & \vb u \geq \bm 0
\end{array},
\end{equation}
where $\vb q_i = (\vb x_i^\T, 1)^\T$, and $\delta = 10^{-8}$ is a small constant to ensure the determinant is nonzero for sets of points that lie in a subspace of dimension lower than the ambient dimension (e.g., completely flat sheets). The equation describing the ellipsoid can then be calculated using the optimum, $\vb u^*$, as $(\vb x - \vb c)^\T \vb Q (\vb x - \vb c) = 1$, where $\vb Q = \frac{1}{3} (\vb X \mathrm{diag}(\vb u^*) \vb X^\T - \vb c \vb c^\T)^{-1}$, $\vb c = \vb X \vb u^*$, and $\vb X = [\vb x_1, \ldots, \vb x_N]$. As used in our previous work \cite{silmore2021}, the orientational covariance matrix describes the spread of the unit normals of triangles across the sheet about a mean orientation. The mean orientation of a sheet lives on the unit sphere like all of the unit normals and is consequently calculated via the weighted Fréchet mean \cite{pennec2006},
\begin{equation}
\bar{\vb n} = \argmin_{\vb x \in S^2} \sum_{i \in \{\bigtriangleup\}} w_i \dist(\vb x, \vu n_i)^2,
\end{equation}
where the sum is over all triangles of the mesh, $\vu n_i$ is the normal of triangle $i$, $w_i$ is the area of triangle $i$, and $\dist : (\vb x, \vb y) \mapsto \arccos(\vb x \vdot \vb y)$ is the Riemannian distance between two points on the sphere. The (weighted) orientational covariance matrix, then, is given by:
\begin{equation}
\bar{\vb C} = \frac{\sum_j w_j}{\qty( \sum_j w_j )^2 - \sum_j w_j^2} \sum_{i \in \{\bigtriangleup\}} w_i \log_{\bar{\vb n}}(\vu n_i) \log_{\bar{\vb n}}(\vu n_i)^\T
\end{equation}
where $\log_{\bar{\vb n}}: S^2 \to T_{\bar{\vb n}} S^2$ is the logarithmic map and maps points on the sphere to the tangent plane at $\bar{\vb n}$. The orientational covariance matrix can be further characterized using its two eigenvalues, $\lambda_1$ and $\lambda_2$, or their sum and difference. Physically, the sum represents the total degree of crumpling, and the magnitude of the difference relative to the sum indicates the anisotropy of crumpling (see Figure 8 of ref. \cite{silmore2021} for an illustration).

Figure \ref{fig:ellipsoid_multi} shows the three semiaxis lengths of minimum-volume bounding ellipsoids over time for several different values of $S$ and $k_B T / \kappa$, both spanning two orders of magnitude. For the stiffest sheets (larger $S$) featured in Figure \ref{fig:ellipsoid_multi} (the bottom two rows), the stochastic flipping discussed in the previous section is immediately evident as peaks in the semiaxis lengths. During flipping events, the largest semiaxis length does not change appreciably, but the two smallest semiaxis lengths tend to approach each other, moreso for the stiffest sheet. Geometrically, this means stiffer sheets are more ``cigar''-like during flipping events, whereas softer sheets are more flattened. It is also important to note that the basal length of the smallest semiaxis increases with temperature, which reflects the fact that greater thermal fluctuations serve to increase the effective thickness about the flat state in the flow-vorticity plane. For $S = 3.08 \times 10^{-5}$ (the top row of Figure \ref{fig:ellipsoid_multi}), the sheet is crumpled and continuously tumbles, as evidenced by the qualitatively and quantitatively different behavior of the semiaxis lengths. All three semiaxis lengths are much closer together for all time compared to the sheets of larger $S$, indicating a bounding ellipsoid that is ``closer'' to a sphere (i.e., less anisotropic), and there are no discrete flipping events that occur. Additionally, one can see that the time required for the sheet to adopt the crumpled conformation from the flat initial condition at time $\dot \gamma t = 0$ generally decreases with increasing temperature, indicating that thermal fluctuations help induce crumpling in sheets that are strongly sheared.

Figure \ref{fig:volume_covariance} shows the time-averaged volume of minimum-volume bounding ellipsoids as well as the time-averaged eigenvalues of $\bar{\vb C}$ as a function of $S$ and $k_B T / \kappa$. First, in all three panels of Figure \ref{fig:volume_covariance}, the transition to crumpling and tumbling around $S \approx 2 \times 10^{-4}$ is quite evident as is the smoothing or ``rounding'' of the transition as temperature increases. One can see in panel A of Figure \ref{fig:volume_covariance} that for values of $S$ above the dynamical crumpling transition, the time-averaged bounding volume generally increases with $S$ at constant temperature and increases with temperature at constant $S$. In light of the stochastic flipping data in Figure \ref{fig:flipping_times_multi} and analysis above, this behavior can largely be attributed to greater flipping frequencies with increasing $S$ and $k_B T / \kappa$ and the larger associated bounding volumes of sheets during flips. Interestingly, for the largest value of $S$ featured in Figure \ref{fig:volume_covariance}A, the volume \textit{decreases} when approaching $k_B T / \kappa = 1$, which further indicates that sufficiently strong thermal fluctuations in the presence of shear can induce more compact conformations, as is the case with the classic \textit{equilibrium} ``crumpling transition'' (distinct from the dynamic crumpling transition in $S$ discussed in this work) in tethered-membranes without self-avoidance \cite{nelson2004}. Whether there is a relationship between this non-monotonic geometric behavior and the \textit{equilibrium} crumpling transition is an interesting question that may be explored in future work. 

Panels B and C of Figure \ref{fig:volume_covariance} show the time-averaged sum and the normalized difference of the eigenvalues of the orientational covariance matrix, $\bar{\vb C}$. While the sum indicates the total degree of variance of the normals across the sheet about the mean orientation (and hence total degree of crumpling), the normalized difference represents the degree to the which the sheet deforms anisotropically. For example, a normalized eigenvalue difference of 1.0 corresponds to a sheet creased in one direction and exhibiting zero curvature in the direction orthogonal to the crease. In panel B, one can see, as expected, that the total degree of crumpling increases monotonically as $S$ decreases for all temperatures and increases sharply around the dynamical crumpling/continuous tumbling transition. The total degree of crumpling also increases with temperature and can likely be attributed to the amplitude of the thermal fluctuations themselves.

In panel C, one can see that larger thermal fluctuations lead to more isotropically crumpled conformations for all values of $S$ above the dynamical crumpling/continuous tumbling transition. Somewhat counterintuitively, the relative degree of anisotropy in deformation \textit{decreases} with $S$ at the highest temperatures but \textit{increases} with $S$ at the lowest temperatures. This behavior may be explained as follows. At high dimensionless temperatures, sheets are flipping quite frequently due to strong thermal fluctuations, and larger directional shear forces relative to bending forces (as $S$ \textit{decreases}) should induce more anisotropy in the deformations during flipping. At low dimensionless temperatures, sheets are flipping relatively infrequently and already deforming with large anisotropy as they fold over in the flow (see movies in the Supporting Information). Greater bending rigidity as $S$ increases (regardless of shear strength), then, should promote more anisotropic folding, which is in line with our conclusions for athermal sheets \cite{silmore2021}. In other words, at large $k_B T / \kappa$, conformational behavior is dominated mostly by the interplay of thermal fluctuations and shear flow, whereas for small $k_B T / \kappa$, conformational behavior is dominated mostly by the interplay of bending rigidity and shear flow. The ratio $k_B T / \kappa$, after all, indicates which energy scale is more relevant.

\section{Rheological Properties}

For a dilute suspension of force-free rigid particles, Batchelor \cite{batchelor1970a} showed that the total stress is given by:
\begin{equation}
\bm \Sigma = -\langle p \rangle \vb I + 2\eta \vb E^\infty + n \langle \tilde{\bm \Sigma} \rangle,
\label{eq:suspension_stress}
\end{equation}
where $p$ is the pressure, $\vb E^\infty = (\vb L + \vb L^\T)/2$ is the imposed rate-of-strain tensor, $n$ is the number concentration of particles, and $\tilde{\bm \Sigma}$ is the stresslet (denoted with $\tilde{\bm \Sigma}$ to avoid confusion with $S$, as it is usually denoted in the literature). The stresslet represents the symmetric part of the first moment of the force distribution on a particle \cite{kim2005, guazzelli2011}. Importantly, all of the angle brackets here represent volume averages, and in an abuse of notation, we conflate the meaning of these angle brackets with the others throughout this paper that represent time averages under the assumption that they should be equal (i.e., ergodicity holds).

The calculation of the stress (and consequently viscosity) from immersed boundary simulations of flexible materials, such as those conducted in this work, is challenging. The Kirkwood-Riseman methodology \cite{kirkwood1948} is often used to model the viscosity of polymer chains, but there are certain mathematical issues (viz., singularities) associated with its use that are often underappreciated \cite{zwanzig1968}. As such, we chose to use the minimum-volume bounding ellipsoids discussed above to estimate the transport properties of sheets. Namely, the stresslet, $\tilde{\bm \Sigma}$, of the bounding ellipsoid at each snapshot was calculated numerically using the formulas found in Kim and Karrila \cite{kim2005} with rotational velocities set to those dictated by Jeffery's equations \cite{jeffery1922} for force- and torque-free ellipsoids in Stokes flow. If the sheet were rigid or if the beads comprising the sheet were force-free, then the energy dissipated by the bounding ellipsoid and its viscosity would be rigorous upper bounds to those of the sheet \cite{kim2005, nir1975}. However, given that the beads are not force-free due to bending and stretching forces, the energy dissipated by the ellipsoid should be considered an approximate upper bound to that of the sheet, capturing the dominant hydrodynamic contributions to the stress due to changes in sheet conformation.

\begin{figure}[ht]
\centering
\includegraphics[width=\linewidth]{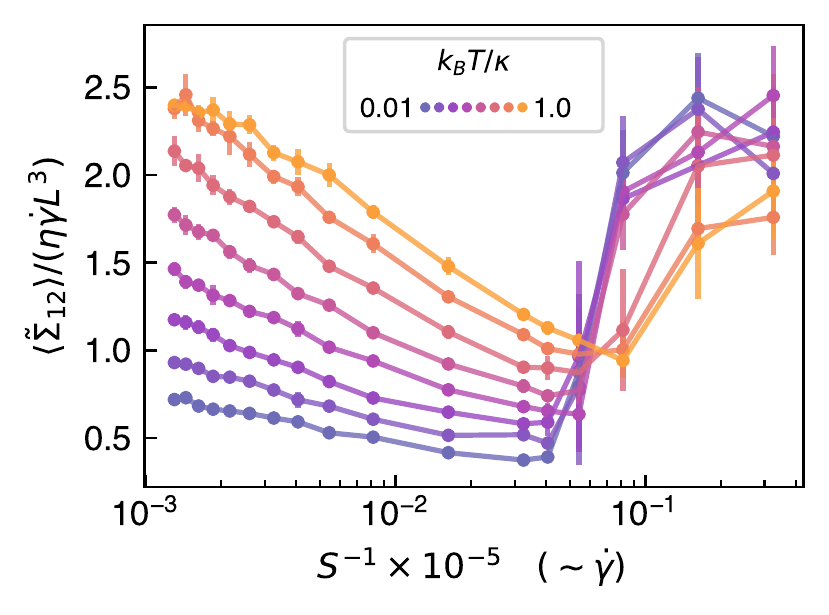}
\caption{Time-averaged dimensionless stresslet contribution to the viscosity (estimated with minimum-volume bounding ellipsoids) as a function of dimensionless bending rigidity, $S$, and dimensionless temperature, $k_B T / \kappa$. Error bars represent two standard errors of the mean between the independent runs, and lines are drawn to guide the eye.}
\label{fig:viscosity}
\end{figure}

Figure \ref{fig:viscosity} shows the time-averaged off-diagonal ``flow-gradient'' entry of the minimum-volume bounding ellipsoid stresslet. Physically, the increase in the effective viscosity of a dilute suspension of sheets should grow linearly with this quantity as described in equation \ref{eq:suspension_stress}. In Figure \ref{fig:viscosity}, this dimensionless stress is plotted against $S^{-1}$ instead of $S$ in order to make it more amenable to typical rheological interpretation since $S^{-1}$ is proportional to the shear rate for \textit{fixed} bending rigidity. For values of $S$ above the dynamical crumpling/continuously tumbling transition, the stresslet contribution to the viscosity increases with temperature and decreases with shear rate. Past the transition, though, the viscosity contribution begins to increase again. This non-monotonic behavior implies that dilute suspensions of semiflexible sheets should be shear-thinning up to to a dimensionless shear rate around $S^{-1} \approx 5 \times 10^{3}$, at which point the suspension exhibits crumpling-induced shear-thickening. Interestingly, dilute suspensions of graphene oxide, a material with a bending rigidity of $\kappa / k_B T \approx 1$ at room temperature \cite{poulin2016}, have indeed been found to exhibit peculiar behavior of shear-thinning followed by shear-thickening \cite{zhang2018}. In fact, Figure 7 in ref. \cite{zhang2018} also demonstrates temperature-dependent viscosity effects that are qualitatively very similar to those depicted in Figure \ref{fig:viscosity} of this work. Although Zhang et al. \cite{zhang2018} do not claim such behavior is due to conformational changes, we believe it is a strong possibility that merits further investigation.

For most values of $S$ examined, sheets do not tumble like rigid platelets, and there is a complicated balance between geometrical conformations and time spent flipping that affects the observed stresslet contribution to the viscosity (see Appendix B). However, as shown in Figure \ref{fig:viscosity_master}, an empirical power-law fit of the form $\tilde \Sigma_{12} / (\eta \dot \gamma L^3) = c_1 S^{c_2} (k_B T / \kappa)^{c_3}$, where $c_1 = 8.55$, $c_2 = 0.236$, and $c_3 = 0.263$, seems to collapse the data before the dynamical crumpling transition well. These constants were calculated via a weighted least-squares fit to the power-law functional form using run-averaged data with $S > 2.5 \times 10^{-4}$ and errors equal to the standard errors of the mean among the independent runs. Assuming independent, Gaussian-distributed errors and a uniform prior, Monte Carlo sampling yielded the following 95\% equal-tailed credible intervals for the fitting parameters: $c_1 \in [8.45, 8.67]$, $c_2 \in [0.2338, 0.2379]$, $c_3 \in [0.2616, 0.2647]$. This scaling with $S^{0.236}$ is largely attributable to the size of the ``U-turn'' radius of the sheet as it flips, which theory predicts should scale as $S^{1/4}$ \cite{harasim2013} (see Appendix B). In terms of shear rate, it is expected, then, that the contribution to viscosity (in the relevant regime of $S$) should exhibit a power-law scaling exponent with respect to shear rate of $-0.236$, or, equivalently, an exponent value of $0.764$ for the power-law fluid model \cite{bird1987}.

\begin{figure}[ht]
\centering
\includegraphics[width=\linewidth]{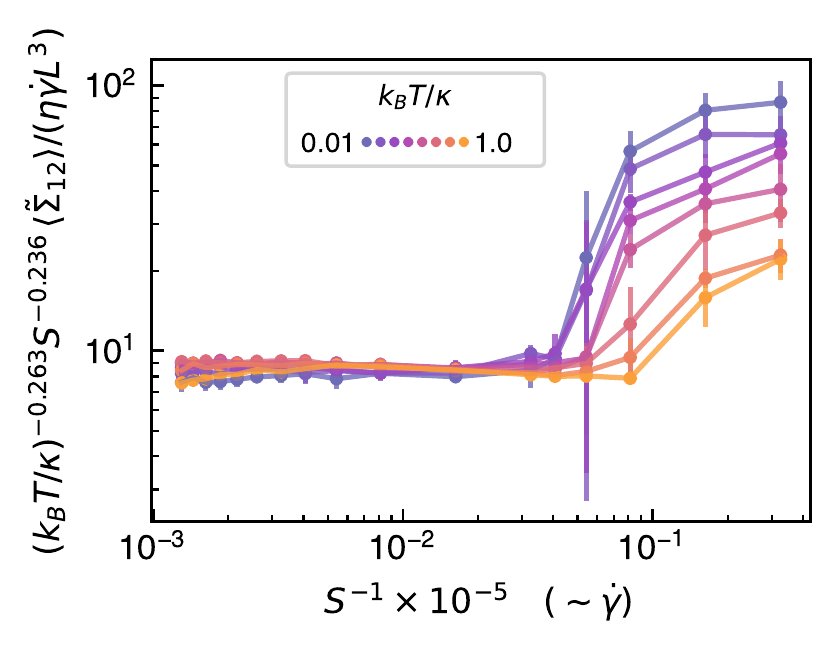}
\caption{Scaled, time-averaged dimensionless stresslet contribution to the viscosity (see Figure \ref{fig:viscosity}) as a function of dimensionless bending rigidity, $S$, and dimensionless temperature, $k_B T / \kappa$. Error bars represent two standard errors of the mean between the independent runs, and lines are drawn to guide the eye.}
\label{fig:viscosity_master}
\end{figure}

\begin{figure*}[ht]
\centering
\includegraphics[width=\linewidth]{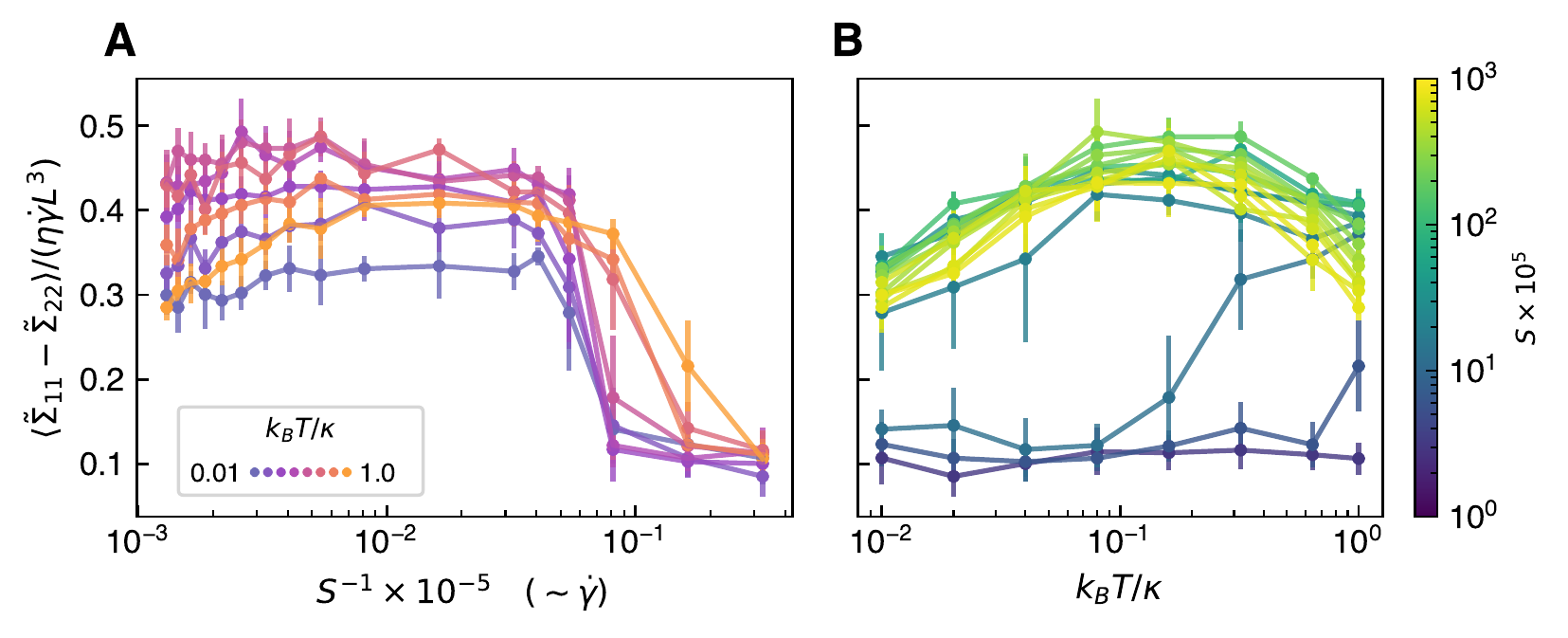}
\caption{Time-averaged dimensionless first normal stress difference (estimated with minimum-volume bounding ellipsoids) as a function of dimensionless bending rigidity, $S$, and dimensionless temperature, $k_B T / \kappa$. Panels \textbf{A} and \textbf{B} show the same data but visualized with $S^{-1}$ or $k_B T / \kappa$ on the x-axis, respectively. Error bars represent two standard errors of the mean between the independent runs, and lines are drawn to guide the eye.}
\label{fig:normal_stress_combined}
\end{figure*}

Regarding the zero-shear viscosity of dilute suspensions of sheets, a power-law fluid model is not appropriate. In fact, it is challenging to calculate such a quantity with the computational model of an asymptotically thin sheet studied in this work. Consider the following: for small $k_B T / \kappa$, as shear rate approaches zero, neither the shear flow nor thermal fluctuations are perturbing the sheet much away from a flat conformation. One may expect, then, that $\kappa$ becomes irrelevant and that the P\'eclet number, $\mathrm{Pe} = S^{-1} (k_B T / \kappa)^{-1}$, becomes the relevant dimensionless group. As shear rate approaches zero, the orientations of the effectively rigid sheets would be uniformly distributed, and the the suspension viscosity would be given by the orientationally averaged stresslet, which, in turn, depends significantly on the thickness of the sheets. For large $k_B T / \kappa$, the zero-shear viscosity would be related to equilibrium sheet conformations, which may be dependent on the chemistry of particular experimental systems of interest \cite{li2020}.

In addition to viscosity, first normal stress differences can be calculated using the stresslet data from the minimum-volume bounding ellipsoids. Figure \ref{fig:normal_stress_combined} shows the time-averaged first normal stress difference as a function of $S$ and $k_B T / \kappa$. As a function of $S$, the first normal stress difference is greatest for values of $S$ above the dynamical crumpling transition (low shear rates) and decreases rapidly for continuously tumbling, crumpled sheets past the transition. This behavior is consistent with the orientational covariance data presented in Figure \ref{fig:volume_covariance} in that first normal stresses differences are most prominent for highly anisotropic particles but sheets beyond the dynamical crumpling transition are more isotropically crumpled. Suspensions of spherical particles, after all, do not exhibit any normal stress differences. Panel B of Figure \ref{fig:normal_stress_combined} shows that, as a function of temperature, stochastically flipping sheets with $S$ values below the dynamical crumpling transition exhibit a local maximum of normal stress between $k_B T / \kappa = 0.1$ to $0.3$. This non-monotonic behavior in temperature is likely due to the fact that both average volume and degree of anisotropic crumpling affect the first normal stress difference. Again, from Figure \ref{fig:volume_covariance}, one can see opposite trends with respect to temperature for the total degree of crumpling (as measured by the sum of the eigenvalues) and the relative anisotropy of deformation (as measured by the normalized difference of the eigenvalues). That is, qualitatively, the balance between these two quantities leads to the observed local maximum.

\section{Conclusions}

Numerical immersed boundary simulations of semiflexible sheets subjected to thermal fluctuations and ambient simple shear flow were conducted over a range of values for dimensionless bending rigidity, $S = \kappa / (\pi \eta \dot \gamma L^3)$, and dimensionless temperature, $k_B T / \kappa$. A dynamical transition from stochastic flipping to indefinitely crumpled and continuously tumbling sheets was identified at a value of $S \approx 2 \times 10^{-4}$, which is consistent with the transition to chaotic crumpling and tumbling found in athermal sheets \cite{silmore2021}. As temperature was increased, this dynamical transition become more gradual and ``rounded''. Scaling arguments similar to those found in the tumbling polymer literature were used to characterize the mean time between flips ($\dot \gamma \Delta t_\mathrm{diff} \sim \left( k_B T / \kappa \right)^{-1/3} S^{-1/3}$) for stochastically flipping sheets that were stiff relative to both shear strength and thermal fluctuations. Additionally, a 1D first passage time model was constructed and successfully applied to explain the scaling of the variance of the flipping time distributions ($\mathrm{Var}[\dot \gamma \Delta t_\mathrm{flip}] \sim \left( k_B T / \kappa \right)^{-2/3} S^{-2/3}$).

Geometric and conformational behavior was quantified via the calculation of minimum-volume bounding ellipsoids as well as orientational covariance matrices describing the spread of normals across the sheet about a mean orientation. Stresslets were calculated for minimum-volume bounding ellipsoids over time in order to estimate the viscosity and first normal stress difference for dilute suspensions of semiflexible sheets.
In particular, up to the dynamical crumping/continuously tumbling transition, suspensions were shear-thinning, and at higher shear rates beyond the dynamical crumpling transition, suspensions were shear-thickening.
%In particular, up to the dynamical crumping/continuously tumbling transition, suspensions were shear-thinning, and a scaling argument was successfully used to construct a master curve for the stresslet contribution to the viscosity in this regime ($\tilde \Sigma_{12} / (\eta \dot \gamma L^3)  \sim (k_B T / \kappa)^{1/3} S^{1/4}$). At higher shear rates beyond the dynamical crumpling transition, suspensions were shear-thickening.
We also observed nonzero first normal stress differences that exhibited a local maximum in temperature and decreased sharply with increasing shear rate beyond the dynamical crumpling transition due to less anisotropy in the dynamically crumpled/continuously tumbling state.

Although the effects of thermal-fluctuation-induced bending rigidity renormalization were mentioned in the introduction, they were not explicitly considered in this work. In particular, the ``bare'' dimensionless bending rigidity was used in all analyses even though thermal fluctuations are known to induce a length-scale-dependent renormalized bending rigidity. With differently sized sheets, instead of scaling with $\kappa$ and $L^3$, one may predict that $S$ should scale like $S \sim \kappa q_\mathrm{th}^{\eta_\kappa} / (\eta \dot \gamma L^{3-\eta_\kappa})$ for large Föppl-von Kármán (FvK) numbers (i.e., sheets with bending modes much softer than stretching modes), where $\eta_\kappa \approx 0.8$  and $q_\mathrm{th} = \sqrt{3 k_B T k / (8 \sqrt{3} \pi \kappa^2)}$ is an inverse thermal length scale \cite{bowick2017, nelson2004}.
Future work that more concretely examines these effects and, namely, whether the bending rigidity in all of the scaling analyses and in the governing dimensionless parameter $S$ could simply be replaced by its renormalized counterpart would be especially valuable.

Between the viscosity and the first normal stress difference, it is clear that the dynamical and conformational behavior of colloidal 2D materials explored in this work contributes to a rich variety of non-Newtonian rheological properties. Importantly, this behavior can be  exploited to design responsive soft materials and appropriately tune solution processing protocols for 2D materials depending on the application. We believe the fundamental advances of this work will greatly inform future theoretical work on sheet dynamics as well as experimental design.

\begin{acknowledgments}
K.S.S. was supported by the U.S. Department of Energy, Office of Science, Office of Advanced Scientific Computing Research, Department of Energy Computational Science Graduate Fellowship under Award Number DE-FG02-97ER25308. M.S.S. was supported by the Department of Energy, Office of Science, Basic Energy Sciences under grant No. DE-FG02-08ER46488 Mod 0008. J.W.S. was supported by NSF Career Award No. CBET-1554398. The authors would like to thank G. McKinley, P. Doyle, and N. Fakhri for helpful discussions.

This report was prepared as an account of work sponsored by an agency of the United States Government. Neither the United States Government nor any agency thereof, nor any of their employees, makes any warranty, express or implied, or assumes any legal liability or responsibility for the accuracy, completeness, or usefulness of any information, apparatus, product, or process disclosed, or represents that its use would not infringe privately owned rights. Reference herein to any specific commercial product, process, or service by trade name, trademark, manufacturer, or otherwise does not necessarily constitute or imply its endorsement, recommendation, or favoring by the United States Government or any agency thereof. The views and opinions of authors expressed herein do not necessarily state or reflect those of the United States Government or any agency thereof.
\end{acknowledgments}

\appendix
\section{First passage time model}

\begin{figure*}[ht]
\centering
\includegraphics{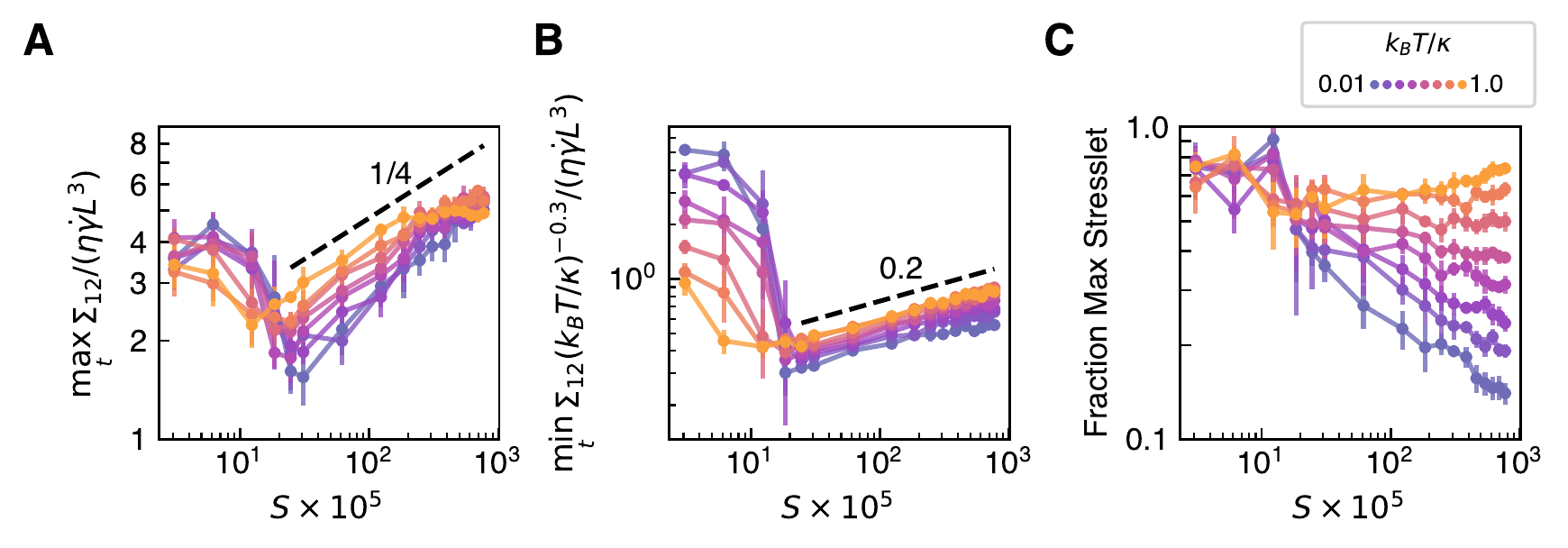}
\caption{\textbf{A)} Maximum instantaneous dimensionless stresslet contribution to the viscosity. \textbf{B)} Minimum instantaneous dimensionless stresslet contribution to the viscosity scaled by $(k_B T / \kappa)^{-0.3}$ in order to reflect the theoretically predicted dependence of height fluctuations on dimensionless temperature. \textbf{C)} Fraction of time spent with instantaneous dimensionless stresslet values in the top 75\% of values attained. Error bars in all plots represent two standard errors of the mean between the independent runs, and lines are drawn to guide the eye.}
\label{fig:stresslet_analysis}
\end{figure*}

Consider a Brownian particle with diffusivity $D_r$ initially located in the center of the domain $[-\theta_c, \theta_c]$ with a reflecting boundary condition on the left and an absorbing boundary condition on the right. The evolution of the probability distribution of the particle over time is governed by the following Fokker-Planck equation:
\begin{equation}
\begin{split}
\partial_{\tilde t} p &= \partial_{xx} p \\
\partial_x p(\tilde t, -\theta_c) = p(\tilde t, \theta_c) &= 0 \\
p(0, x) &= \delta(x)
\end{split}
\end{equation}
where $\tilde t = D_r t$. An infinite series solution for $p$ is given by:
\begin{equation}
p(\tilde t, x) = \frac{1}{\theta_c} \sum_{n=1}^\infty \sin(\lambda_n \theta_c) \exp(-\lambda_n^2 \tilde t) \sin(\lambda_n(\theta_c - x)),
\end{equation}
where
\begin{equation}
\lambda_n = \frac{(2n - 1) \pi}{4 \theta_c}.
\end{equation}
The survival probability (i.e., the probability the particle has remained in the domain at time $\tilde t$) is defined as $S_p(\tilde t) = \int_{-\theta_c}^{\theta_c} p(\tilde t, x) \dd{x}$, and the associated first passage time distribution is defined as $f(\tilde t) = -\mathrm{d}S_p / \mathrm{d}\tilde t$. Analytically, $f(\tilde t)$ can also be represented by an infinite series:
\begin{equation}
f(\tilde t) = \frac{1}{\theta_c} \sum_{n=1}^\infty \lambda_n \sin(\lambda_n \theta_c) \exp(-\lambda_n^2 \tilde t) \left[ 1 - \cos(2\lambda_n \theta_c) \right].
\end{equation}
In terms of $\theta_c$ and $D_r$, the mean and variance of the first passage time are found to be
\begin{equation}
\mathbb E[f] = \frac{3}{2}\frac{\theta_c^2}{D_r}, \qquad \mathrm{Var}[f] = \frac{5}{2}\frac{\theta_c^4}{D_r^2}.
\end{equation}
This form of the mean first passage time further justifies the diffusive time scale that was balanced against the advective time scale in the scaling analysis of stochastic flipping times.

\section{Stresslet analysis}

Figure \ref{fig:stresslet_analysis} shows the maximum and minimum instantaneous dimensionless stresslet contributions to the viscosity as well as the fraction of time spent with instantaneous stresslet values in the top 75\% of those attained (i.e., time that can largely be attributed to flipping or crumpling in the flow), all averaged over the six independent runs. The minimum instantaneous stresslet values in panel B are scaled by the dimensionless temperature to a certain power, as explained below.

In panel A of Figure \ref{fig:stresslet_analysis}, it can be seen that the maximum instantaneous off-diagonal stresslet values attained are not a strong function of dimensionless temperature. For ``intermediate'' values of $S$ that represent sheets that are not approaching infinite stiffness (i.e., most of the range of values examined beyond that of the dynamical crumpling transition), this behavior can be explained with geometric reasoning and a force balance. Following the arguments of Harasim et al. \cite{harasim2013} for understanding the ``U-turn'' radius of a tumbling polymer, balancing the hydrodynamic force on a ``U-turn'' in a flipping sheet with the bending moment of the ``U-turn'' yields the following scaling for the radius of curvature:
\begin{equation}
h \sim \left( \frac{\kappa L}{\eta \dot \gamma} \right)^{1/4}.
\end{equation}
Furthermore, given that the stresslet of an ellipsoid scales with its volume and assuming the size of the sheet in the flow and vorticity directions scales with $L$ (supported by the data in Figure \ref{fig:volume_covariance}), then
\begin{equation}
\max_t \frac{\tilde \Sigma_{12}}{\eta \dot \gamma L^3}
\sim \frac{h}{L} \sim \frac{1}{L} \left( \frac{\kappa L}{\eta \dot \gamma} \right)^{1/4} \sim S^{1/4}.
\end{equation}
This scaling with $S$ is depicted in panel A and agrees reasonably well with the data. One can see deviations for the highest dimensionless temperature, which can likely be attributed  the inapplicability of the ``U-turn'' flipping model, as larger thermal fluctuations induce frequent flipping associated with more variable deformation.

Panel B shows the minimum instantaneous off-diagonal stresslet values attained, which can be attributed physically to thermal height fluctuations of flat sheets in the flow-vorticity plane (for those sheets with $S$ values above the dynamical crumpling transition). By a similar argument, given that the bounding ellipsoid stresslet scales like $hL^2$ and using known results \cite{bowick2017, nelson2004} for the average height fluctuations of a thermalized tethered membrane:
\begin{equation}
\begin{split}
\min_t \frac{\tilde \Sigma_{12}}{\eta \dot \gamma L^3} &\sim \frac{h}{L} \sim \frac{1}{L} \sqrt{\frac{k_B T}{L^2 \kappa_r(q) q^4}} \\
&\sim \sqrt{\frac{k_B T}{\kappa (q / q_\mathrm{th})^{-\eta_\kappa} }} \\
&\sim \sqrt{\frac{k_B T}{\kappa}} \left( \frac{k_B T Y L^2}{\kappa^2} \right)^{-\eta_\kappa / 4} \\
&\sim \left( \frac{k_B T}{\kappa} \right)^{1/2 - \eta_\kappa / 4} S^{\eta_\kappa / 4},
\end{split}
\end{equation}
where it is assumed the relevant wavelength scales as $q \sim 1/L$, and the weak scaling with $S$ in the last line follows from the fact that the interbead bond strength employed and described in the Model and Methods section scales with the shear flow strength. Such a scaling argument seems to explain the data in panel B quite well. Some of the discrepancies in panels A and B between the data and the proposed scaling arguments can likely be attributed to the inexact scaling of the stresslet with the height fluctuations / U-turn radius as well as deviations in the geometrical conformations from those underlying the relatively simple physical arguments employed.

Panel C of Figure \ref{fig:stresslet_analysis} shows the fraction of time spent with instantaneous stresslet values in the top 75\% of those attained, which is one measure of the fraction of time spent flipping or at least ``disturbing'' the ambient shear flow. While the dependence on $S$ and dimensionless temperature over the full range of $S$ examined is complicated, it is this quantity when multiplied by the maximum instantaneous stresslet values in panel A that results in the empirical power-law scaling in $S$ and $k_B T / \kappa$ seen in Figure \ref{fig:viscosity_master}. The minimum instantaneous stresslet values are approximately an order of magnitude smaller than the time-averaged stresslet values. Thus, the weak dependence on $S$ (or, equivalently, FvK number) seen in panel B due to the choice of harmonic bond potential strength negligibly impacts the observed time-averaged stresslet contribution to the viscosity. That is, the observed viscosity scaling should be expected to hold for a given material with a \textit{constant} FvK number over a relevant range of dimensionless shear rates ($S^{-1}$).

\bibliography{extracted.bib}

\begin{thebibliography}{79}%
\makeatletter
\providecommand \@ifxundefined [1]{%
 \@ifx{#1\undefined}
}%
\providecommand \@ifnum [1]{%
 \ifnum #1\expandafter \@firstoftwo
 \else \expandafter \@secondoftwo
 \fi
}%
\providecommand \@ifx [1]{%
 \ifx #1\expandafter \@firstoftwo
 \else \expandafter \@secondoftwo
 \fi
}%
\providecommand \natexlab [1]{#1}%
\providecommand \enquote  [1]{``#1''}%
\providecommand \bibnamefont  [1]{#1}%
\providecommand \bibfnamefont [1]{#1}%
\providecommand \citenamefont [1]{#1}%
\providecommand \href@noop [0]{\@secondoftwo}%
\providecommand \href [0]{\begingroup \@sanitize@url \@href}%
\providecommand \@href[1]{\@@startlink{#1}\@@href}%
\providecommand \@@href[1]{\endgroup#1\@@endlink}%
\providecommand \@sanitize@url [0]{\catcode `\\12\catcode `\$12\catcode
  `\&12\catcode `\#12\catcode `\^12\catcode `\_12\catcode `\%12\relax}%
\providecommand \@@startlink[1]{}%
\providecommand \@@endlink[0]{}%
\providecommand \url  [0]{\begingroup\@sanitize@url \@url }%
\providecommand \@url [1]{\endgroup\@href {#1}{\urlprefix }}%
\providecommand \urlprefix  [0]{URL }%
\providecommand \Eprint [0]{\href }%
\providecommand \doibase [0]{https://doi.org/}%
\providecommand \selectlanguage [0]{\@gobble}%
\providecommand \bibinfo  [0]{\@secondoftwo}%
\providecommand \bibfield  [0]{\@secondoftwo}%
\providecommand \translation [1]{[#1]}%
\providecommand \BibitemOpen [0]{}%
\providecommand \bibitemStop [0]{}%
\providecommand \bibitemNoStop [0]{.\EOS\space}%
\providecommand \EOS [0]{\spacefactor3000\relax}%
\providecommand \BibitemShut  [1]{\csname bibitem#1\endcsname}%
\let\auto@bib@innerbib\@empty
%</preamble>
\bibitem [{\citenamefont {Stafford}\ \emph {et~al.}(2018)\citenamefont
  {Stafford}, \citenamefont {Patapas}, \citenamefont {Uzo}, \citenamefont
  {Matar},\ and\ \citenamefont {Petit}}]{stafford2018}%
  \BibitemOpen
  \bibfield  {author} {\bibinfo {author} {\bibfnamefont {J.}~\bibnamefont
  {Stafford}}, \bibinfo {author} {\bibfnamefont {A.}~\bibnamefont {Patapas}},
  \bibinfo {author} {\bibfnamefont {N.}~\bibnamefont {Uzo}}, \bibinfo {author}
  {\bibfnamefont {O.~K.}\ \bibnamefont {Matar}},\ and\ \bibinfo {author}
  {\bibfnamefont {C.}~\bibnamefont {Petit}},\ }\bibfield  {title} {\bibinfo
  {title} {Towards scale-up of graphene production via nonoxidizing liquid
  exfoliation methods},\ }\href {https://doi.org/10.1002/aic.16174} {\bibfield
  {journal} {\bibinfo  {journal} {AIChE Journal}\ }\textbf {\bibinfo {volume}
  {64}},\ \bibinfo {pages} {3246} (\bibinfo {year} {2018})}\BibitemShut
  {NoStop}%
\bibitem [{\citenamefont {Ambrosi}\ and\ \citenamefont
  {Pumera}(2018)}]{ambrosi2018}%
  \BibitemOpen
  \bibfield  {author} {\bibinfo {author} {\bibfnamefont {A.}~\bibnamefont
  {Ambrosi}}\ and\ \bibinfo {author} {\bibfnamefont {M.}~\bibnamefont
  {Pumera}},\ }\bibfield  {title} {\bibinfo {title} {Exfoliation of layered
  materials using electrochemistry},\ }\href
  {https://doi.org/10.1039/C7CS00811B} {\bibfield  {journal} {\bibinfo
  {journal} {Chem. Soc. Rev.}\ }\textbf {\bibinfo {volume} {47}},\ \bibinfo
  {pages} {7213} (\bibinfo {year} {2018})}\BibitemShut {NoStop}%
\bibitem [{\citenamefont {Coleman}\ \emph {et~al.}(2011)\citenamefont
  {Coleman}, \citenamefont {Lotya}, \citenamefont {O'Neill}, \citenamefont
  {Bergin}, \citenamefont {King}, \citenamefont {Khan}, \citenamefont {Young},
  \citenamefont {Gaucher}, \citenamefont {De}, \citenamefont {Smith},
  \citenamefont {Shvets}, \citenamefont {Arora}, \citenamefont {Stanton},
  \citenamefont {Kim}, \citenamefont {Lee}, \citenamefont {Kim}, \citenamefont
  {Duesberg}, \citenamefont {Hallam}, \citenamefont {Boland}, \citenamefont
  {Wang}, \citenamefont {Donegan}, \citenamefont {Grunlan}, \citenamefont
  {Moriarty}, \citenamefont {Shmeliov}, \citenamefont {Nicholls}, \citenamefont
  {Perkins}, \citenamefont {Grieveson}, \citenamefont {Theuwissen},
  \citenamefont {McComb}, \citenamefont {Nellist},\ and\ \citenamefont
  {Nicolosi}}]{coleman2011}%
  \BibitemOpen
  \bibfield  {author} {\bibinfo {author} {\bibfnamefont {J.~N.}\ \bibnamefont
  {Coleman}}, \bibinfo {author} {\bibfnamefont {M.}~\bibnamefont {Lotya}},
  \bibinfo {author} {\bibfnamefont {A.}~\bibnamefont {O'Neill}}, \bibinfo
  {author} {\bibfnamefont {S.~D.}\ \bibnamefont {Bergin}}, \bibinfo {author}
  {\bibfnamefont {P.~J.}\ \bibnamefont {King}}, \bibinfo {author}
  {\bibfnamefont {U.}~\bibnamefont {Khan}}, \bibinfo {author} {\bibfnamefont
  {K.}~\bibnamefont {Young}}, \bibinfo {author} {\bibfnamefont
  {A.}~\bibnamefont {Gaucher}}, \bibinfo {author} {\bibfnamefont
  {S.}~\bibnamefont {De}}, \bibinfo {author} {\bibfnamefont {R.~J.}\
  \bibnamefont {Smith}}, \bibinfo {author} {\bibfnamefont {I.~V.}\ \bibnamefont
  {Shvets}}, \bibinfo {author} {\bibfnamefont {S.~K.}\ \bibnamefont {Arora}},
  \bibinfo {author} {\bibfnamefont {G.}~\bibnamefont {Stanton}}, \bibinfo
  {author} {\bibfnamefont {H.-Y.}\ \bibnamefont {Kim}}, \bibinfo {author}
  {\bibfnamefont {K.}~\bibnamefont {Lee}}, \bibinfo {author} {\bibfnamefont
  {G.~T.}\ \bibnamefont {Kim}}, \bibinfo {author} {\bibfnamefont {G.~S.}\
  \bibnamefont {Duesberg}}, \bibinfo {author} {\bibfnamefont {T.}~\bibnamefont
  {Hallam}}, \bibinfo {author} {\bibfnamefont {J.~J.}\ \bibnamefont {Boland}},
  \bibinfo {author} {\bibfnamefont {J.~J.}\ \bibnamefont {Wang}}, \bibinfo
  {author} {\bibfnamefont {J.~F.}\ \bibnamefont {Donegan}}, \bibinfo {author}
  {\bibfnamefont {J.~C.}\ \bibnamefont {Grunlan}}, \bibinfo {author}
  {\bibfnamefont {G.}~\bibnamefont {Moriarty}}, \bibinfo {author}
  {\bibfnamefont {A.}~\bibnamefont {Shmeliov}}, \bibinfo {author}
  {\bibfnamefont {R.~J.}\ \bibnamefont {Nicholls}}, \bibinfo {author}
  {\bibfnamefont {J.~M.}\ \bibnamefont {Perkins}}, \bibinfo {author}
  {\bibfnamefont {E.~M.}\ \bibnamefont {Grieveson}}, \bibinfo {author}
  {\bibfnamefont {K.}~\bibnamefont {Theuwissen}}, \bibinfo {author}
  {\bibfnamefont {D.~W.}\ \bibnamefont {McComb}}, \bibinfo {author}
  {\bibfnamefont {P.~D.}\ \bibnamefont {Nellist}},\ and\ \bibinfo {author}
  {\bibfnamefont {V.}~\bibnamefont {Nicolosi}},\ }\bibfield  {title} {\bibinfo
  {title} {Two-{{Dimensional Nanosheets Produced}} by {{Liquid Exfoliation}} of
  {{Layered Materials}}},\ }\href {https://doi.org/10.1126/science.1194975}
  {\bibfield  {journal} {\bibinfo  {journal} {Science}\ }\textbf {\bibinfo
  {volume} {331}},\ \bibinfo {pages} {568} (\bibinfo {year}
  {2011})}\BibitemShut {NoStop}%
\bibitem [{\citenamefont {Nicolosi}\ \emph {et~al.}(2013)\citenamefont
  {Nicolosi}, \citenamefont {Chhowalla}, \citenamefont {Kanatzidis},
  \citenamefont {Strano},\ and\ \citenamefont {Coleman}}]{nicolosi2013}%
  \BibitemOpen
  \bibfield  {author} {\bibinfo {author} {\bibfnamefont {V.}~\bibnamefont
  {Nicolosi}}, \bibinfo {author} {\bibfnamefont {M.}~\bibnamefont {Chhowalla}},
  \bibinfo {author} {\bibfnamefont {M.~G.}\ \bibnamefont {Kanatzidis}},
  \bibinfo {author} {\bibfnamefont {M.~S.}\ \bibnamefont {Strano}},\ and\
  \bibinfo {author} {\bibfnamefont {J.~N.}\ \bibnamefont {Coleman}},\
  }\bibfield  {title} {\bibinfo {title} {Liquid {{Exfoliation}} of {{Layered
  Materials}}},\ }\href {https://doi.org/10.1126/science.1226419} {\bibfield
  {journal} {\bibinfo  {journal} {Science}\ }\textbf {\bibinfo {volume}
  {340}},\ \bibinfo {pages} {1226419} (\bibinfo {year} {2013})}\BibitemShut
  {NoStop}%
\bibitem [{\citenamefont {Varrla}\ \emph {et~al.}(2015)\citenamefont {Varrla},
  \citenamefont {Backes}, \citenamefont {Paton}, \citenamefont {Harvey},
  \citenamefont {Gholamvand}, \citenamefont {McCauley},\ and\ \citenamefont
  {Coleman}}]{varrla2015}%
  \BibitemOpen
  \bibfield  {author} {\bibinfo {author} {\bibfnamefont {E.}~\bibnamefont
  {Varrla}}, \bibinfo {author} {\bibfnamefont {C.}~\bibnamefont {Backes}},
  \bibinfo {author} {\bibfnamefont {K.~R.}\ \bibnamefont {Paton}}, \bibinfo
  {author} {\bibfnamefont {A.}~\bibnamefont {Harvey}}, \bibinfo {author}
  {\bibfnamefont {Z.}~\bibnamefont {Gholamvand}}, \bibinfo {author}
  {\bibfnamefont {J.}~\bibnamefont {McCauley}},\ and\ \bibinfo {author}
  {\bibfnamefont {J.~N.}\ \bibnamefont {Coleman}},\ }\bibfield  {title}
  {\bibinfo {title} {Large-{{Scale Production}} of {{Size}}-{{Controlled MoS2
  Nanosheets}} by {{Shear Exfoliation}}},\ }\href
  {https://doi.org/10.1021/cm5044864} {\bibfield  {journal} {\bibinfo
  {journal} {Chem. Mater.}\ }\textbf {\bibinfo {volume} {27}},\ \bibinfo
  {pages} {1129} (\bibinfo {year} {2015})}\BibitemShut {NoStop}%
\bibitem [{\citenamefont {Mallory}\ \emph {et~al.}(2015)\citenamefont
  {Mallory}, \citenamefont {Valeriani},\ and\ \citenamefont
  {Cacciuto}}]{mallory2015}%
  \BibitemOpen
  \bibfield  {author} {\bibinfo {author} {\bibfnamefont {S.~A.}\ \bibnamefont
  {Mallory}}, \bibinfo {author} {\bibfnamefont {C.}~\bibnamefont {Valeriani}},\
  and\ \bibinfo {author} {\bibfnamefont {A.}~\bibnamefont {Cacciuto}},\
  }\bibfield  {title} {\bibinfo {title} {Anomalous dynamics of an elastic
  membrane in an active fluid},\ }\href
  {https://doi.org/10.1103/PhysRevE.92.012314} {\bibfield  {journal} {\bibinfo
  {journal} {Phys. Rev. E}\ }\textbf {\bibinfo {volume} {92}},\ \bibinfo
  {pages} {012314} (\bibinfo {year} {2015})}\BibitemShut {NoStop}%
\bibitem [{\citenamefont {Gibaud}\ \emph {et~al.}(2017)\citenamefont {Gibaud},
  \citenamefont {Kaplan}, \citenamefont {Sharma}, \citenamefont {Zakhary},
  \citenamefont {Ward}, \citenamefont {Oldenbourg}, \citenamefont {Meyer},
  \citenamefont {Kamien}, \citenamefont {Powers},\ and\ \citenamefont
  {Dogic}}]{gibaud2017a}%
  \BibitemOpen
  \bibfield  {author} {\bibinfo {author} {\bibfnamefont {T.}~\bibnamefont
  {Gibaud}}, \bibinfo {author} {\bibfnamefont {C.~N.}\ \bibnamefont {Kaplan}},
  \bibinfo {author} {\bibfnamefont {P.}~\bibnamefont {Sharma}}, \bibinfo
  {author} {\bibfnamefont {M.~J.}\ \bibnamefont {Zakhary}}, \bibinfo {author}
  {\bibfnamefont {A.}~\bibnamefont {Ward}}, \bibinfo {author} {\bibfnamefont
  {R.}~\bibnamefont {Oldenbourg}}, \bibinfo {author} {\bibfnamefont {R.~B.}\
  \bibnamefont {Meyer}}, \bibinfo {author} {\bibfnamefont {R.~D.}\ \bibnamefont
  {Kamien}}, \bibinfo {author} {\bibfnamefont {T.~R.}\ \bibnamefont {Powers}},\
  and\ \bibinfo {author} {\bibfnamefont {Z.}~\bibnamefont {Dogic}},\ }\bibfield
   {title} {\bibinfo {title} {Achiral symmetry breaking and positive
  {{Gaussian}} modulus lead to scalloped colloidal membranes},\ }\href
  {https://doi.org/10.1073/pnas.1617043114} {\bibfield  {journal} {\bibinfo
  {journal} {Proc. Natl. Acad. Sci. U.S.A.}\ }\textbf {\bibinfo {volume}
  {114}},\ \bibinfo {pages} {E3376} (\bibinfo {year} {2017})}\BibitemShut
  {NoStop}%
\bibitem [{\citenamefont {Ding}\ \emph {et~al.}(2017)\citenamefont {Ding},
  \citenamefont {Liu}, \citenamefont {Zeng}, \citenamefont {Xia}, \citenamefont
  {Wells}, \citenamefont {Nieh},\ and\ \citenamefont {Sun}}]{ding2017}%
  \BibitemOpen
  \bibfield  {author} {\bibinfo {author} {\bibfnamefont {F.}~\bibnamefont
  {Ding}}, \bibinfo {author} {\bibfnamefont {J.}~\bibnamefont {Liu}}, \bibinfo
  {author} {\bibfnamefont {S.}~\bibnamefont {Zeng}}, \bibinfo {author}
  {\bibfnamefont {Y.}~\bibnamefont {Xia}}, \bibinfo {author} {\bibfnamefont
  {K.~M.}\ \bibnamefont {Wells}}, \bibinfo {author} {\bibfnamefont {M.-P.}\
  \bibnamefont {Nieh}},\ and\ \bibinfo {author} {\bibfnamefont
  {L.}~\bibnamefont {Sun}},\ }\bibfield  {title} {\bibinfo {title} {Biomimetic
  nanocoatings with exceptional mechanical, barrier, and flame-retardant
  properties from large-scale one-step coassembly},\ }\href@noop {} {\bibfield
  {journal} {\bibinfo  {journal} {Science Advances}\ }\textbf {\bibinfo
  {volume} {3}},\ \bibinfo {pages} {e1701212} (\bibinfo {year}
  {2017})}\BibitemShut {NoStop}%
\bibitem [{\citenamefont {Tardani}\ \emph {et~al.}(2018)\citenamefont
  {Tardani}, \citenamefont {Neri}, \citenamefont {Zakri}, \citenamefont
  {Kellay}, \citenamefont {Colin},\ and\ \citenamefont
  {Poulin}}]{tardani2018a}%
  \BibitemOpen
  \bibfield  {author} {\bibinfo {author} {\bibfnamefont {F.}~\bibnamefont
  {Tardani}}, \bibinfo {author} {\bibfnamefont {W.}~\bibnamefont {Neri}},
  \bibinfo {author} {\bibfnamefont {C.}~\bibnamefont {Zakri}}, \bibinfo
  {author} {\bibfnamefont {H.}~\bibnamefont {Kellay}}, \bibinfo {author}
  {\bibfnamefont {A.}~\bibnamefont {Colin}},\ and\ \bibinfo {author}
  {\bibfnamefont {P.}~\bibnamefont {Poulin}},\ }\bibfield  {title} {\bibinfo
  {title} {Shear {{Rheology Control}} of {{Wrinkles}} and {{Patterns}} in
  {{Graphene Oxide Films}}},\ }\href
  {https://doi.org/10.1021/acs.langmuir.7b04281} {\bibfield  {journal}
  {\bibinfo  {journal} {Langmuir}\ }\textbf {\bibinfo {volume} {34}},\ \bibinfo
  {pages} {2996} (\bibinfo {year} {2018})}\BibitemShut {NoStop}%
\bibitem [{\citenamefont {Davidson}\ \emph {et~al.}(2018)\citenamefont
  {Davidson}, \citenamefont {Penisson}, \citenamefont {Constantin},\ and\
  \citenamefont {Gabriel}}]{davidson2018a}%
  \BibitemOpen
  \bibfield  {author} {\bibinfo {author} {\bibfnamefont {P.}~\bibnamefont
  {Davidson}}, \bibinfo {author} {\bibfnamefont {C.}~\bibnamefont {Penisson}},
  \bibinfo {author} {\bibfnamefont {D.}~\bibnamefont {Constantin}},\ and\
  \bibinfo {author} {\bibfnamefont {J.-C.~P.}\ \bibnamefont {Gabriel}},\
  }\bibfield  {title} {\bibinfo {title} {Isotropic, nematic, and lamellar
  phases in colloidal suspensions of nanosheets},\ }\href
  {https://doi.org/10.1073/pnas.1802692115} {\bibfield  {journal} {\bibinfo
  {journal} {Proc. Natl. Acad. Sci. U.S.A.}\ ,\ \bibinfo {pages} {201802692}}
  (\bibinfo {year} {2018})}\BibitemShut {NoStop}%
\bibitem [{\citenamefont {Laskar}\ \emph {et~al.}(2018)\citenamefont {Laskar},
  \citenamefont {Shklyaev},\ and\ \citenamefont {Balazs}}]{laskar2018}%
  \BibitemOpen
  \bibfield  {author} {\bibinfo {author} {\bibfnamefont {A.}~\bibnamefont
  {Laskar}}, \bibinfo {author} {\bibfnamefont {O.~E.}\ \bibnamefont
  {Shklyaev}},\ and\ \bibinfo {author} {\bibfnamefont {A.~C.}\ \bibnamefont
  {Balazs}},\ }\bibfield  {title} {\bibinfo {title} {Designing self-propelled,
  chemically active sheets: {{Wrappers}}, flappers, and creepers},\ }\href
  {https://doi.org/10.1126/sciadv.aav1745} {\bibfield  {journal} {\bibinfo
  {journal} {Science Advances}\ }\textbf {\bibinfo {volume} {4}},\ \bibinfo
  {pages} {eaav1745} (\bibinfo {year} {2018})}\BibitemShut {NoStop}%
\bibitem [{\citenamefont {Kamal}\ \emph {et~al.}(2020)\citenamefont {Kamal},
  \citenamefont {Gravelle},\ and\ \citenamefont {Botto}}]{kamal2020}%
  \BibitemOpen
  \bibfield  {author} {\bibinfo {author} {\bibfnamefont {C.}~\bibnamefont
  {Kamal}}, \bibinfo {author} {\bibfnamefont {S.}~\bibnamefont {Gravelle}},\
  and\ \bibinfo {author} {\bibfnamefont {L.}~\bibnamefont {Botto}},\ }\bibfield
   {title} {\bibinfo {title} {Hydrodynamic slip can align thin nanoplatelets in
  shear flow},\ }\href {https://doi.org/10.1038/s41467-020-15939-w} {\bibfield
  {journal} {\bibinfo  {journal} {Nature Communications}\ }\textbf {\bibinfo
  {volume} {11}},\ \bibinfo {pages} {2425} (\bibinfo {year}
  {2020})}\BibitemShut {NoStop}%
\bibitem [{\citenamefont {Pezzulla}\ \emph {et~al.}(2020)\citenamefont
  {Pezzulla}, \citenamefont {Strong}, \citenamefont {Gallaire},\ and\
  \citenamefont {Reis}}]{pezzulla2020}%
  \BibitemOpen
  \bibfield  {author} {\bibinfo {author} {\bibfnamefont {M.}~\bibnamefont
  {Pezzulla}}, \bibinfo {author} {\bibfnamefont {E.~F.}\ \bibnamefont
  {Strong}}, \bibinfo {author} {\bibfnamefont {F.}~\bibnamefont {Gallaire}},\
  and\ \bibinfo {author} {\bibfnamefont {P.~M.}\ \bibnamefont {Reis}},\
  }\bibfield  {title} {\bibinfo {title} {Deformation of porous flexible strip
  in low and moderate {{Reynolds}} number flows},\ }\href
  {https://doi.org/10.1103/PhysRevFluids.5.084103} {\bibfield  {journal}
  {\bibinfo  {journal} {Phys. Rev. Fluids}\ }\textbf {\bibinfo {volume} {5}},\
  \bibinfo {pages} {084103} (\bibinfo {year} {2020})}\BibitemShut {NoStop}%
\bibitem [{\citenamefont {Klotz}\ \emph {et~al.}(2020)\citenamefont {Klotz},
  \citenamefont {Soh},\ and\ \citenamefont {Doyle}}]{klotz2020}%
  \BibitemOpen
  \bibfield  {author} {\bibinfo {author} {\bibfnamefont {A.~R.}\ \bibnamefont
  {Klotz}}, \bibinfo {author} {\bibfnamefont {B.~W.}\ \bibnamefont {Soh}},\
  and\ \bibinfo {author} {\bibfnamefont {P.~S.}\ \bibnamefont {Doyle}},\
  }\bibfield  {title} {\bibinfo {title} {Equilibrium structure and deformation
  response of {{2D}} kinetoplast sheets},\ }\href
  {https://doi.org/10.1073/pnas.1911088116} {\bibfield  {journal} {\bibinfo
  {journal} {Proc. Natl. Acad. Sci. U.S.A.}\ }\textbf {\bibinfo {volume}
  {117}},\ \bibinfo {pages} {121} (\bibinfo {year} {2020})}\BibitemShut
  {NoStop}%
\bibitem [{\citenamefont {Liu}\ \emph {et~al.}(2018)\citenamefont {Liu},
  \citenamefont {Liu}, \citenamefont {Kozawa}, \citenamefont {Dong},
  \citenamefont {Yang}, \citenamefont {Koman}, \citenamefont {Saccone},
  \citenamefont {Wang}, \citenamefont {Son}, \citenamefont {Wong},\ and\
  \citenamefont {Strano}}]{liu2018}%
  \BibitemOpen
  \bibfield  {author} {\bibinfo {author} {\bibfnamefont {P.}~\bibnamefont
  {Liu}}, \bibinfo {author} {\bibfnamefont {A.~T.}\ \bibnamefont {Liu}},
  \bibinfo {author} {\bibfnamefont {D.}~\bibnamefont {Kozawa}}, \bibinfo
  {author} {\bibfnamefont {J.}~\bibnamefont {Dong}}, \bibinfo {author}
  {\bibfnamefont {J.~F.}\ \bibnamefont {Yang}}, \bibinfo {author}
  {\bibfnamefont {V.~B.}\ \bibnamefont {Koman}}, \bibinfo {author}
  {\bibfnamefont {M.}~\bibnamefont {Saccone}}, \bibinfo {author} {\bibfnamefont
  {S.}~\bibnamefont {Wang}}, \bibinfo {author} {\bibfnamefont {Y.}~\bibnamefont
  {Son}}, \bibinfo {author} {\bibfnamefont {M.~H.}\ \bibnamefont {Wong}},\ and\
  \bibinfo {author} {\bibfnamefont {M.~S.}\ \bibnamefont {Strano}},\ }\bibfield
   {title} {\bibinfo {title} {Autoperforation of {{2D}} materials for
  generating two-terminal memristive {{Janus}} particles},\ }\href
  {https://doi.org/10.1038/s41563-018-0197-z} {\bibfield  {journal} {\bibinfo
  {journal} {Nature Materials}\ }\textbf {\bibinfo {volume} {17}},\ \bibinfo
  {pages} {1005} (\bibinfo {year} {2018})}\BibitemShut {NoStop}%
\bibitem [{\citenamefont {Jeffery}(1922)}]{jeffery1922}%
  \BibitemOpen
  \bibfield  {author} {\bibinfo {author} {\bibfnamefont {G.~B.}\ \bibnamefont
  {Jeffery}},\ }\bibfield  {title} {\bibinfo {title} {The {{Motion}} of
  {{Ellipsoidal Particles Immersed}} in a {{Viscous Fluid}}},\ }\href
  {https://doi.org/10.1098/rspa.1922.0078} {\bibfield  {journal} {\bibinfo
  {journal} {Proc. R. Soc. London, Ser. A}\ }\textbf {\bibinfo {volume}
  {102}},\ \bibinfo {pages} {161} (\bibinfo {year} {1922})}\BibitemShut
  {NoStop}%
\bibitem [{\citenamefont {Hinch}\ and\ \citenamefont
  {Leal}(1972)}]{hinch1972a}%
  \BibitemOpen
  \bibfield  {author} {\bibinfo {author} {\bibfnamefont {E.~J.}\ \bibnamefont
  {Hinch}}\ and\ \bibinfo {author} {\bibfnamefont {L.~G.}\ \bibnamefont
  {Leal}},\ }\bibfield  {title} {\bibinfo {title} {The effect of {{Brownian}}
  motion on the rheological properties of a suspension of non-spherical
  particles},\ }\href {https://doi.org/10.1017/S002211207200271X} {\bibfield
  {journal} {\bibinfo  {journal} {Journal of Fluid Mechanics}\ }\textbf
  {\bibinfo {volume} {52}},\ \bibinfo {pages} {683} (\bibinfo {year}
  {1972})}\BibitemShut {NoStop}%
\bibitem [{\citenamefont {Hinch}\ and\ \citenamefont {Leal}(1979)}]{hinch1979}%
  \BibitemOpen
  \bibfield  {author} {\bibinfo {author} {\bibfnamefont {E.~J.}\ \bibnamefont
  {Hinch}}\ and\ \bibinfo {author} {\bibfnamefont {L.~G.}\ \bibnamefont
  {Leal}},\ }\bibfield  {title} {\bibinfo {title} {Rotation of small
  non-axisymmetric particles in a simple shear flow},\ }\href
  {https://doi.org/10.1017/S002211207900077X} {\bibfield  {journal} {\bibinfo
  {journal} {Journal of Fluid Mechanics}\ }\textbf {\bibinfo {volume} {92}},\
  \bibinfo {pages} {591} (\bibinfo {year} {1979})}\BibitemShut {NoStop}%
\bibitem [{\citenamefont {Leal}\ and\ \citenamefont {Hinch}(1971)}]{leal1971}%
  \BibitemOpen
  \bibfield  {author} {\bibinfo {author} {\bibfnamefont {L.~G.}\ \bibnamefont
  {Leal}}\ and\ \bibinfo {author} {\bibfnamefont {E.~J.}\ \bibnamefont
  {Hinch}},\ }\bibfield  {title} {\bibinfo {title} {The effect of weak
  {{Brownian}} rotations on particles in shear flow},\ }\href
  {https://doi.org/10.1017/S0022112071000788} {\bibfield  {journal} {\bibinfo
  {journal} {Journal of Fluid Mechanics}\ }\textbf {\bibinfo {volume} {46}},\
  \bibinfo {pages} {685} (\bibinfo {year} {1971})}\BibitemShut {NoStop}%
\bibitem [{\citenamefont {Batchelor}(1970{\natexlab{a}})}]{batchelor1970}%
  \BibitemOpen
  \bibfield  {author} {\bibinfo {author} {\bibfnamefont {G.~K.}\ \bibnamefont
  {Batchelor}},\ }\bibfield  {title} {\bibinfo {title} {Slender-body theory for
  particles of arbitrary cross-section in {{Stokes}} flow},\ }\href
  {https://doi.org/10.1017/S002211207000191X} {\bibfield  {journal} {\bibinfo
  {journal} {J. Fluid Mech.}\ }\textbf {\bibinfo {volume} {44}},\ \bibinfo
  {pages} {419} (\bibinfo {year} {1970}{\natexlab{a}})}\BibitemShut {NoStop}%
\bibitem [{\citenamefont {Batchelor}(1970{\natexlab{b}})}]{batchelor1970a}%
  \BibitemOpen
  \bibfield  {author} {\bibinfo {author} {\bibfnamefont {G.~K.}\ \bibnamefont
  {Batchelor}},\ }\bibfield  {title} {\bibinfo {title} {The stress system in a
  suspension of force-free particles},\ }\href
  {https://doi.org/10.1017/S0022112070000745} {\bibfield  {journal} {\bibinfo
  {journal} {Journal of Fluid Mechanics}\ }\textbf {\bibinfo {volume} {41}},\
  \bibinfo {pages} {545} (\bibinfo {year} {1970}{\natexlab{b}})}\BibitemShut
  {NoStop}%
\bibitem [{\citenamefont {Bird}\ \emph {et~al.}(1987)\citenamefont {Bird},
  \citenamefont {Armstrong},\ and\ \citenamefont {Hassager}}]{bird1987}%
  \BibitemOpen
  \bibfield  {author} {\bibinfo {author} {\bibfnamefont {R.~B.}\ \bibnamefont
  {Bird}}, \bibinfo {author} {\bibfnamefont {R.~C.}\ \bibnamefont
  {Armstrong}},\ and\ \bibinfo {author} {\bibfnamefont {O.}~\bibnamefont
  {Hassager}},\ }\href@noop {} {\emph {\bibinfo {title} {Dynamics of Polymeric
  Liquids}}},\ \bibinfo {edition} {2nd}\ ed.\ (\bibinfo  {publisher}
  {{Wiley}},\ \bibinfo {address} {{New York}},\ \bibinfo {year}
  {1987})\BibitemShut {NoStop}%
\bibitem [{\citenamefont {Silmore}\ \emph {et~al.}(2021)\citenamefont
  {Silmore}, \citenamefont {Strano},\ and\ \citenamefont {Swan}}]{silmore2021}%
  \BibitemOpen
  \bibfield  {author} {\bibinfo {author} {\bibfnamefont {K.~S.}\ \bibnamefont
  {Silmore}}, \bibinfo {author} {\bibfnamefont {M.~S.}\ \bibnamefont
  {Strano}},\ and\ \bibinfo {author} {\bibfnamefont {J.~W.}\ \bibnamefont
  {Swan}},\ }\bibfield  {title} {\bibinfo {title} {Buckling, crumpling, and
  tumbling of semiflexible sheets in simple shear flow},\ }\href
  {https://doi.org/10.1039/D0SM02184A} {\bibfield  {journal} {\bibinfo
  {journal} {Soft Matter}\ }\textbf {\bibinfo {volume} {17}},\ \bibinfo {pages}
  {4707} (\bibinfo {year} {2021})}\BibitemShut {NoStop}%
\bibitem [{\citenamefont {Nelson}\ \emph {et~al.}(2004)\citenamefont {Nelson},
  \citenamefont {Piran},\ and\ \citenamefont {Weinberg}}]{nelson2004}%
  \BibitemOpen
  \bibfield  {author} {\bibinfo {author} {\bibfnamefont {D.}~\bibnamefont
  {Nelson}}, \bibinfo {author} {\bibfnamefont {T.}~\bibnamefont {Piran}},\ and\
  \bibinfo {author} {\bibfnamefont {S.}~\bibnamefont {Weinberg}},\ }\href
  {https://doi.org/10.1142/5473} {\emph {\bibinfo {title} {Statistical
  {{Mechanics}} of {{Membranes}} and {{Surfaces}}}}},\ \bibinfo {edition}
  {2nd}\ ed.\ (\bibinfo  {publisher} {{World Scientific}},\ \bibinfo {year}
  {2004})\BibitemShut {NoStop}%
\bibitem [{\citenamefont {Aronovitz}\ and\ \citenamefont
  {Lubensky}(1988)}]{aronovitz1988}%
  \BibitemOpen
  \bibfield  {author} {\bibinfo {author} {\bibfnamefont {J.~A.}\ \bibnamefont
  {Aronovitz}}\ and\ \bibinfo {author} {\bibfnamefont {T.~C.}\ \bibnamefont
  {Lubensky}},\ }\bibfield  {title} {\bibinfo {title} {Fluctuations of {{Solid
  Membranes}}},\ }\href {https://doi.org/10.1103/PhysRevLett.60.2634}
  {\bibfield  {journal} {\bibinfo  {journal} {Phys. Rev. Lett.}\ }\textbf
  {\bibinfo {volume} {60}},\ \bibinfo {pages} {2634} (\bibinfo {year}
  {1988})}\BibitemShut {NoStop}%
\bibitem [{\citenamefont {Frey}\ and\ \citenamefont {Nelson}(1991)}]{frey1991}%
  \BibitemOpen
  \bibfield  {author} {\bibinfo {author} {\bibfnamefont {E.}~\bibnamefont
  {Frey}}\ and\ \bibinfo {author} {\bibfnamefont {D.~L.}\ \bibnamefont
  {Nelson}},\ }\bibfield  {title} {\bibinfo {title} {Dynamics of flat membranes
  and flickering in red blood cells},\ }\href
  {https://doi.org/10.1051/jp1:1991238} {\bibfield  {journal} {\bibinfo
  {journal} {J. Phys. I}\ }\textbf {\bibinfo {volume} {1}},\ \bibinfo {pages}
  {1715} (\bibinfo {year} {1991})}\BibitemShut {NoStop}%
\bibitem [{\citenamefont {Kantor}\ \emph {et~al.}(1986)\citenamefont {Kantor},
  \citenamefont {Kardar},\ and\ \citenamefont {Nelson}}]{kantor1986}%
  \BibitemOpen
  \bibfield  {author} {\bibinfo {author} {\bibfnamefont {Y.}~\bibnamefont
  {Kantor}}, \bibinfo {author} {\bibfnamefont {M.}~\bibnamefont {Kardar}},\
  and\ \bibinfo {author} {\bibfnamefont {D.~R.}\ \bibnamefont {Nelson}},\
  }\bibfield  {title} {\bibinfo {title} {Statistical {{Mechanics}} of
  {{Tethered Surfaces}}},\ }\href {https://doi.org/10.1103/PhysRevLett.57.791}
  {\bibfield  {journal} {\bibinfo  {journal} {Phys. Rev. Lett.}\ }\textbf
  {\bibinfo {volume} {57}},\ \bibinfo {pages} {791} (\bibinfo {year}
  {1986})}\BibitemShut {NoStop}%
\bibitem [{\citenamefont {Kardar}\ and\ \citenamefont
  {Nelson}(1988)}]{kardar1988}%
  \BibitemOpen
  \bibfield  {author} {\bibinfo {author} {\bibfnamefont {M.}~\bibnamefont
  {Kardar}}\ and\ \bibinfo {author} {\bibfnamefont {D.~R.}\ \bibnamefont
  {Nelson}},\ }\bibfield  {title} {\bibinfo {title} {Statistical mechanics of
  self-avoiding tethered manifolds},\ }\href
  {https://doi.org/10.1103/PhysRevA.38.966} {\bibfield  {journal} {\bibinfo
  {journal} {Phys. Rev. A}\ }\textbf {\bibinfo {volume} {38}},\ \bibinfo
  {pages} {966} (\bibinfo {year} {1988})}\BibitemShut {NoStop}%
\bibitem [{\citenamefont {Nelson}\ and\ \citenamefont
  {Peliti}(1987)}]{nelson1987}%
  \BibitemOpen
  \bibfield  {author} {\bibinfo {author} {\bibfnamefont {D.}~\bibnamefont
  {Nelson}}\ and\ \bibinfo {author} {\bibfnamefont {L.}~\bibnamefont
  {Peliti}},\ }\bibfield  {title} {\bibinfo {title} {Fluctuations in membranes
  with crystalline and hexatic order},\ }\href
  {https://doi.org/10.1051/jphys:019870048070108500} {\bibfield  {journal}
  {\bibinfo  {journal} {J. Phys. (Paris)}\ }\textbf {\bibinfo {volume} {48}},\
  \bibinfo {pages} {1085} (\bibinfo {year} {1987})}\BibitemShut {NoStop}%
\bibitem [{\citenamefont {Paczuski}\ \emph {et~al.}(1988)\citenamefont
  {Paczuski}, \citenamefont {Kardar},\ and\ \citenamefont
  {Nelson}}]{paczuski1988}%
  \BibitemOpen
  \bibfield  {author} {\bibinfo {author} {\bibfnamefont {M.}~\bibnamefont
  {Paczuski}}, \bibinfo {author} {\bibfnamefont {M.}~\bibnamefont {Kardar}},\
  and\ \bibinfo {author} {\bibfnamefont {D.~R.}\ \bibnamefont {Nelson}},\
  }\bibfield  {title} {\bibinfo {title} {Landau {{Theory}} of the {{Crumpling
  Transition}}},\ }\href {https://doi.org/10.1103/PhysRevLett.60.2638}
  {\bibfield  {journal} {\bibinfo  {journal} {Phys. Rev. Lett.}\ }\textbf
  {\bibinfo {volume} {60}},\ \bibinfo {pages} {2638} (\bibinfo {year}
  {1988})}\BibitemShut {NoStop}%
\bibitem [{\citenamefont {Ko{\v s}mrlj}\ and\ \citenamefont
  {Nelson}(2016)}]{kosmrlj2016}%
  \BibitemOpen
  \bibfield  {author} {\bibinfo {author} {\bibfnamefont {A.}~\bibnamefont
  {Ko{\v s}mrlj}}\ and\ \bibinfo {author} {\bibfnamefont {D.~R.}\ \bibnamefont
  {Nelson}},\ }\bibfield  {title} {\bibinfo {title} {Response of thermalized
  ribbons to pulling and bending},\ }\href
  {https://doi.org/10.1103/PhysRevB.93.125431} {\bibfield  {journal} {\bibinfo
  {journal} {Phys. Rev. B}\ }\textbf {\bibinfo {volume} {93}},\ \bibinfo
  {pages} {125431} (\bibinfo {year} {2016})}\BibitemShut {NoStop}%
\bibitem [{\citenamefont {Abraham}\ \emph {et~al.}(1989)\citenamefont
  {Abraham}, \citenamefont {Rudge},\ and\ \citenamefont
  {Plischke}}]{abraham1989}%
  \BibitemOpen
  \bibfield  {author} {\bibinfo {author} {\bibfnamefont {F.~F.}\ \bibnamefont
  {Abraham}}, \bibinfo {author} {\bibfnamefont {W.~E.}\ \bibnamefont {Rudge}},\
  and\ \bibinfo {author} {\bibfnamefont {M.}~\bibnamefont {Plischke}},\
  }\bibfield  {title} {\bibinfo {title} {Molecular dynamics of tethered
  membranes},\ }\href {https://doi.org/10.1103/PhysRevLett.62.1757} {\bibfield
  {journal} {\bibinfo  {journal} {Phys. Rev. Lett.}\ }\textbf {\bibinfo
  {volume} {62}},\ \bibinfo {pages} {1757} (\bibinfo {year}
  {1989})}\BibitemShut {NoStop}%
\bibitem [{\citenamefont {Abraham}\ and\ \citenamefont
  {Nelson}(1990)}]{abraham1990}%
  \BibitemOpen
  \bibfield  {author} {\bibinfo {author} {\bibfnamefont {F.~F.}\ \bibnamefont
  {Abraham}}\ and\ \bibinfo {author} {\bibfnamefont {D.~R.}\ \bibnamefont
  {Nelson}},\ }\bibfield  {title} {\bibinfo {title} {Diffraction from
  {{Polymerized Membranes}}},\ }\href
  {https://doi.org/10.1126/science.249.4967.393} {\bibfield  {journal}
  {\bibinfo  {journal} {Science}\ }\textbf {\bibinfo {volume} {249}},\ \bibinfo
  {pages} {393} (\bibinfo {year} {1990})}\BibitemShut {NoStop}%
\bibitem [{\citenamefont {Bowick}\ \emph {et~al.}(2001)\citenamefont {Bowick},
  \citenamefont {Cacciuto}, \citenamefont {Thorleifsson},\ and\ \citenamefont
  {Travesset}}]{bowick2001}%
  \BibitemOpen
  \bibfield  {author} {\bibinfo {author} {\bibfnamefont {M.}~\bibnamefont
  {Bowick}}, \bibinfo {author} {\bibfnamefont {A.}~\bibnamefont {Cacciuto}},
  \bibinfo {author} {\bibfnamefont {G.}~\bibnamefont {Thorleifsson}},\ and\
  \bibinfo {author} {\bibfnamefont {A.}~\bibnamefont {Travesset}},\ }\bibfield
  {title} {\bibinfo {title} {Universality classes of self-avoiding
  fixed-connectivity membranes},\ }\href
  {https://doi.org/10.1007/s101890170071} {\bibfield  {journal} {\bibinfo
  {journal} {Eur. Phys. J. E}\ }\textbf {\bibinfo {volume} {5}},\ \bibinfo
  {pages} {149} (\bibinfo {year} {2001})}\BibitemShut {NoStop}%
\bibitem [{\citenamefont {Bowick}\ \emph {et~al.}(2017)\citenamefont {Bowick},
  \citenamefont {Ko{\v s}mrlj}, \citenamefont {Nelson},\ and\ \citenamefont
  {Sknepnek}}]{bowick2017}%
  \BibitemOpen
  \bibfield  {author} {\bibinfo {author} {\bibfnamefont {M.~J.}\ \bibnamefont
  {Bowick}}, \bibinfo {author} {\bibfnamefont {A.}~\bibnamefont {Ko{\v
  s}mrlj}}, \bibinfo {author} {\bibfnamefont {D.~R.}\ \bibnamefont {Nelson}},\
  and\ \bibinfo {author} {\bibfnamefont {R.}~\bibnamefont {Sknepnek}},\
  }\bibfield  {title} {\bibinfo {title} {Non-{{Hookean}} statistical mechanics
  of clamped graphene ribbons},\ }\href
  {https://doi.org/10.1103/PhysRevB.95.104109} {\bibfield  {journal} {\bibinfo
  {journal} {Phys. Rev. B}\ }\textbf {\bibinfo {volume} {95}},\ \bibinfo
  {pages} {104109} (\bibinfo {year} {2017})}\BibitemShut {NoStop}%
\bibitem [{\citenamefont {Gompper}\ and\ \citenamefont
  {Kroll}(1997)}]{gompper1997}%
  \BibitemOpen
  \bibfield  {author} {\bibinfo {author} {\bibfnamefont {G.}~\bibnamefont
  {Gompper}}\ and\ \bibinfo {author} {\bibfnamefont {D.~M.}\ \bibnamefont
  {Kroll}},\ }\bibfield  {title} {\bibinfo {title} {Fluctuations of
  polymerized, fluid and hexatic membranes: {{Continuum}} models and
  simulations},\ }\href {https://doi.org/10.1016/S1359-0294(97)80079-9}
  {\bibfield  {journal} {\bibinfo  {journal} {Current Opinion in Colloid \&
  Interface Science}\ }\textbf {\bibinfo {volume} {2}},\ \bibinfo {pages} {373}
  (\bibinfo {year} {1997})}\BibitemShut {NoStop}%
\bibitem [{\citenamefont {Grest}\ and\ \citenamefont
  {Petsche}(1994)}]{grest1994}%
  \BibitemOpen
  \bibfield  {author} {\bibinfo {author} {\bibfnamefont {G.~S.}\ \bibnamefont
  {Grest}}\ and\ \bibinfo {author} {\bibfnamefont {I.~B.}\ \bibnamefont
  {Petsche}},\ }\bibfield  {title} {\bibinfo {title} {Molecular dynamics
  simulations of self-avoiding tethered membranes with attractive interactions:
  {{Search}} for a crumpled phase},\ }\href
  {https://doi.org/10.1103/PhysRevE.50.R1737} {\bibfield  {journal} {\bibinfo
  {journal} {Phys. Rev. E}\ }\textbf {\bibinfo {volume} {50}},\ \bibinfo
  {pages} {R1737} (\bibinfo {year} {1994})}\BibitemShut {NoStop}%
\bibitem [{\citenamefont {Ho}\ and\ \citenamefont
  {Baumg{\"a}rtner}(1989)}]{ho1989}%
  \BibitemOpen
  \bibfield  {author} {\bibinfo {author} {\bibfnamefont {J.-S.}\ \bibnamefont
  {Ho}}\ and\ \bibinfo {author} {\bibfnamefont {A.}~\bibnamefont
  {Baumg{\"a}rtner}},\ }\bibfield  {title} {\bibinfo {title} {Self-avoiding
  tethered membranes},\ }\href {https://doi.org/10.1103/PhysRevLett.63.1324}
  {\bibfield  {journal} {\bibinfo  {journal} {Phys. Rev. Lett.}\ }\textbf
  {\bibinfo {volume} {63}},\ \bibinfo {pages} {1324} (\bibinfo {year}
  {1989})}\BibitemShut {NoStop}%
\bibitem [{\citenamefont {Plischke}\ and\ \citenamefont
  {Boal}(1988)}]{plischke1988}%
  \BibitemOpen
  \bibfield  {author} {\bibinfo {author} {\bibfnamefont {M.}~\bibnamefont
  {Plischke}}\ and\ \bibinfo {author} {\bibfnamefont {D.}~\bibnamefont
  {Boal}},\ }\bibfield  {title} {\bibinfo {title} {Absence of a crumpling
  transition in strongly self-avoiding tethered membranes},\ }\href
  {https://doi.org/10.1103/PhysRevA.38.4943} {\bibfield  {journal} {\bibinfo
  {journal} {Phys. Rev. A}\ }\textbf {\bibinfo {volume} {38}},\ \bibinfo
  {pages} {4943} (\bibinfo {year} {1988})}\BibitemShut {NoStop}%
\bibitem [{\citenamefont {Tr{\"o}ster}(2013)}]{troster2013}%
  \BibitemOpen
  \bibfield  {author} {\bibinfo {author} {\bibfnamefont {A.}~\bibnamefont
  {Tr{\"o}ster}},\ }\bibfield  {title} {\bibinfo {title} {High-precision
  {{Fourier Monte Carlo}} simulation of crystalline membranes},\ }\href
  {https://doi.org/10.1103/PhysRevB.87.104112} {\bibfield  {journal} {\bibinfo
  {journal} {Phys. Rev. B}\ }\textbf {\bibinfo {volume} {87}},\ \bibinfo
  {pages} {104112} (\bibinfo {year} {2013})}\BibitemShut {NoStop}%
\bibitem [{\citenamefont {Spector}\ \emph {et~al.}(1994)\citenamefont
  {Spector}, \citenamefont {Naranjo}, \citenamefont {Chiruvolu},\ and\
  \citenamefont {Zasadzinski}}]{spector1994}%
  \BibitemOpen
  \bibfield  {author} {\bibinfo {author} {\bibfnamefont {M.~S.}\ \bibnamefont
  {Spector}}, \bibinfo {author} {\bibfnamefont {E.}~\bibnamefont {Naranjo}},
  \bibinfo {author} {\bibfnamefont {S.}~\bibnamefont {Chiruvolu}},\ and\
  \bibinfo {author} {\bibfnamefont {J.~A.}\ \bibnamefont {Zasadzinski}},\
  }\bibfield  {title} {\bibinfo {title} {Conformations of a {{Tethered
  Membrane}}: {{Crumpling}} in {{Graphitic Oxide}}?},\ }\href
  {https://doi.org/10.1103/PhysRevLett.73.2867} {\bibfield  {journal} {\bibinfo
   {journal} {Phys. Rev. Lett.}\ }\textbf {\bibinfo {volume} {73}},\ \bibinfo
  {pages} {2867} (\bibinfo {year} {1994})}\BibitemShut {NoStop}%
\bibitem [{\citenamefont {Wen}\ \emph {et~al.}(1992)\citenamefont {Wen},
  \citenamefont {Garland}, \citenamefont {Hwa}, \citenamefont {Kardar},
  \citenamefont {Kokufuta}, \citenamefont {Li}, \citenamefont {Orkisz},\ and\
  \citenamefont {Tanaka}}]{wen1992}%
  \BibitemOpen
  \bibfield  {author} {\bibinfo {author} {\bibfnamefont {X.}~\bibnamefont
  {Wen}}, \bibinfo {author} {\bibfnamefont {C.~W.}\ \bibnamefont {Garland}},
  \bibinfo {author} {\bibfnamefont {T.}~\bibnamefont {Hwa}}, \bibinfo {author}
  {\bibfnamefont {M.}~\bibnamefont {Kardar}}, \bibinfo {author} {\bibfnamefont
  {E.}~\bibnamefont {Kokufuta}}, \bibinfo {author} {\bibfnamefont
  {Y.}~\bibnamefont {Li}}, \bibinfo {author} {\bibfnamefont {M.}~\bibnamefont
  {Orkisz}},\ and\ \bibinfo {author} {\bibfnamefont {T.}~\bibnamefont
  {Tanaka}},\ }\bibfield  {title} {\bibinfo {title} {Crumpled and collapsed
  conformation in graphite oxide membranes},\ }\href
  {https://doi.org/10.1038/355426a0} {\bibfield  {journal} {\bibinfo  {journal}
  {Nature}\ }\textbf {\bibinfo {volume} {355}},\ \bibinfo {pages} {426}
  (\bibinfo {year} {1992})}\BibitemShut {NoStop}%
\bibitem [{\citenamefont {Li}\ \emph {et~al.}(2020)\citenamefont {Li},
  \citenamefont {Wang}, \citenamefont {Meng}, \citenamefont {Wang},
  \citenamefont {Guo}, \citenamefont {Rajendran}, \citenamefont {Gao},
  \citenamefont {Xu},\ and\ \citenamefont {Xu}}]{li2020}%
  \BibitemOpen
  \bibfield  {author} {\bibinfo {author} {\bibfnamefont {P.}~\bibnamefont
  {Li}}, \bibinfo {author} {\bibfnamefont {S.}~\bibnamefont {Wang}}, \bibinfo
  {author} {\bibfnamefont {F.}~\bibnamefont {Meng}}, \bibinfo {author}
  {\bibfnamefont {Y.}~\bibnamefont {Wang}}, \bibinfo {author} {\bibfnamefont
  {F.}~\bibnamefont {Guo}}, \bibinfo {author} {\bibfnamefont {S.}~\bibnamefont
  {Rajendran}}, \bibinfo {author} {\bibfnamefont {C.}~\bibnamefont {Gao}},
  \bibinfo {author} {\bibfnamefont {Z.}~\bibnamefont {Xu}},\ and\ \bibinfo
  {author} {\bibfnamefont {Z.}~\bibnamefont {Xu}},\ }\bibfield  {title}
  {\bibinfo {title} {Conformational {{Scaling Relations}} of
  {{Two}}-{{Dimensional Macromolecular Graphene Oxide}} in {{Solution}}},\
  }\href {https://doi.org/10.1021/acs.macromol.0c01425} {\bibfield  {journal}
  {\bibinfo  {journal} {Macromolecules}\ }\textbf {\bibinfo {volume} {53}},\
  \bibinfo {pages} {10421} (\bibinfo {year} {2020})}\BibitemShut {NoStop}%
\bibitem [{\citenamefont {Xu}\ and\ \citenamefont {Green}(2014)}]{xu2014}%
  \BibitemOpen
  \bibfield  {author} {\bibinfo {author} {\bibfnamefont {Y.}~\bibnamefont
  {Xu}}\ and\ \bibinfo {author} {\bibfnamefont {M.~J.}\ \bibnamefont {Green}},\
  }\bibfield  {title} {\bibinfo {title} {Brownian dynamics simulations of
  nanosheet solutions under shear},\ }\href {https://doi.org/10.1063/1.4884821}
  {\bibfield  {journal} {\bibinfo  {journal} {J. Chem. Phys.}\ }\textbf
  {\bibinfo {volume} {141}},\ \bibinfo {pages} {024905} (\bibinfo {year}
  {2014})}\BibitemShut {NoStop}%
\bibitem [{\citenamefont {Xu}\ and\ \citenamefont {Green}(2015)}]{xu2015}%
  \BibitemOpen
  \bibfield  {author} {\bibinfo {author} {\bibfnamefont {Y.}~\bibnamefont
  {Xu}}\ and\ \bibinfo {author} {\bibfnamefont {M.~J.}\ \bibnamefont {Green}},\
  }\bibfield  {title} {\bibinfo {title} {Brownian dynamics simulation of
  two-dimensional nanosheets under biaxial extensional flow},\ }\href
  {https://doi.org/10.1002/polb.23760} {\bibfield  {journal} {\bibinfo
  {journal} {J. Polym. Sci. Part B: Polym. Phys.}\ }\textbf {\bibinfo {volume}
  {53}},\ \bibinfo {pages} {1247} (\bibinfo {year} {2015})}\BibitemShut
  {NoStop}%
\bibitem [{\citenamefont {Babu}\ and\ \citenamefont {Stark}(2011)}]{babu2011}%
  \BibitemOpen
  \bibfield  {author} {\bibinfo {author} {\bibfnamefont {S.~B.}\ \bibnamefont
  {Babu}}\ and\ \bibinfo {author} {\bibfnamefont {H.}~\bibnamefont {Stark}},\
  }\bibfield  {title} {\bibinfo {title} {Dynamics of semi-flexible tethered
  sheets},\ }\href {https://doi.org/10.1140/epje/i2011-11136-2} {\bibfield
  {journal} {\bibinfo  {journal} {Eur. Phys. J. E}\ }\textbf {\bibinfo {volume}
  {34}},\ \bibinfo {pages} {136} (\bibinfo {year} {2011})}\BibitemShut
  {NoStop}%
\bibitem [{\citenamefont {Dutta}\ and\ \citenamefont
  {Graham}(2017)}]{dutta2017}%
  \BibitemOpen
  \bibfield  {author} {\bibinfo {author} {\bibfnamefont {S.}~\bibnamefont
  {Dutta}}\ and\ \bibinfo {author} {\bibfnamefont {M.~D.}\ \bibnamefont
  {Graham}},\ }\bibfield  {title} {\bibinfo {title} {Dynamics of
  {{Miura}}-patterned foldable sheets in shear flow},\ }\href
  {https://doi.org/10.1039/C6SM02113A} {\bibfield  {journal} {\bibinfo
  {journal} {Soft Matter}\ }\textbf {\bibinfo {volume} {13}},\ \bibinfo {pages}
  {2620} (\bibinfo {year} {2017})}\BibitemShut {NoStop}%
\bibitem [{\citenamefont {Yu}\ and\ \citenamefont {Graham}(2021)}]{yu2021}%
  \BibitemOpen
  \bibfield  {author} {\bibinfo {author} {\bibfnamefont {Y.}~\bibnamefont
  {Yu}}\ and\ \bibinfo {author} {\bibfnamefont {M.~D.}\ \bibnamefont
  {Graham}},\ }\bibfield  {title} {\bibinfo {title} {Coil\textendash
  stretch-like transition of elastic sheets in extensional flows},\ }\href
  {https://doi.org/10.1039/D0SM01630F} {\bibfield  {journal} {\bibinfo
  {journal} {Soft Matter}\ }\textbf {\bibinfo {volume} {17}},\ \bibinfo {pages}
  {543} (\bibinfo {year} {2021})}\BibitemShut {NoStop}%
\bibitem [{\citenamefont {Bian}\ \emph {et~al.}(2020)\citenamefont {Bian},
  \citenamefont {Litvinov},\ and\ \citenamefont {Koumoutsakos}}]{bian2020}%
  \BibitemOpen
  \bibfield  {author} {\bibinfo {author} {\bibfnamefont {X.}~\bibnamefont
  {Bian}}, \bibinfo {author} {\bibfnamefont {S.}~\bibnamefont {Litvinov}},\
  and\ \bibinfo {author} {\bibfnamefont {P.}~\bibnamefont {Koumoutsakos}},\
  }\bibfield  {title} {\bibinfo {title} {Bending models of lipid bilayer
  membranes: {{Spontaneous}} curvature and area-difference elasticity},\ }\href
  {https://doi.org/10.1016/j.cma.2019.112758} {\bibfield  {journal} {\bibinfo
  {journal} {Computer Methods in Applied Mechanics and Engineering}\ }\textbf
  {\bibinfo {volume} {359}},\ \bibinfo {pages} {112758} (\bibinfo {year}
  {2020})}\BibitemShut {NoStop}%
\bibitem [{\citenamefont {Guckenberger}\ and\ \citenamefont
  {Gekle}(2017)}]{guckenberger2017}%
  \BibitemOpen
  \bibfield  {author} {\bibinfo {author} {\bibfnamefont {A.}~\bibnamefont
  {Guckenberger}}\ and\ \bibinfo {author} {\bibfnamefont {S.}~\bibnamefont
  {Gekle}},\ }\bibfield  {title} {\bibinfo {title} {Theory and algorithms to
  compute {{Helfrich}} bending forces: A review},\ }\href
  {https://doi.org/10.1088/1361-648X/aa6313} {\bibfield  {journal} {\bibinfo
  {journal} {J. Phys.: Condens. Matter}\ }\textbf {\bibinfo {volume} {29}},\
  \bibinfo {pages} {203001} (\bibinfo {year} {2017})}\BibitemShut {NoStop}%
\bibitem [{\citenamefont {Gompper}\ and\ \citenamefont
  {Kroll}(1996)}]{gompper1996}%
  \BibitemOpen
  \bibfield  {author} {\bibinfo {author} {\bibfnamefont {G.}~\bibnamefont
  {Gompper}}\ and\ \bibinfo {author} {\bibfnamefont {D.~M.}\ \bibnamefont
  {Kroll}},\ }\bibfield  {title} {\bibinfo {title} {Random {{Surface
  Discretizations}} and the {{Renormalization}} of the {{Bending Rigidity}}},\
  }\href {https://doi.org/10.1051/jp1:1996246} {\bibfield  {journal} {\bibinfo
  {journal} {J. Phys. I}\ }\textbf {\bibinfo {volume} {6}},\ \bibinfo {pages}
  {1305} (\bibinfo {year} {1996})}\BibitemShut {NoStop}%
\bibitem [{\citenamefont {Heyes}\ and\ \citenamefont
  {Melrose}(1993)}]{heyes1993}%
  \BibitemOpen
  \bibfield  {author} {\bibinfo {author} {\bibfnamefont {D.~M.}\ \bibnamefont
  {Heyes}}\ and\ \bibinfo {author} {\bibfnamefont {J.~R.}\ \bibnamefont
  {Melrose}},\ }\bibfield  {title} {\bibinfo {title} {Brownian dynamics
  simulations of model hard-sphere suspensions},\ }\href
  {https://doi.org/10.1016/0377-0257(93)80001-R} {\bibfield  {journal}
  {\bibinfo  {journal} {J. Non-Newtonian Fluid Mech.}\ }\textbf {\bibinfo
  {volume} {46}},\ \bibinfo {pages} {1} (\bibinfo {year} {1993})}\BibitemShut
  {NoStop}%
\bibitem [{\citenamefont {Silmore}\ and\ \citenamefont
  {Swan}(2020)}]{silmore2020}%
  \BibitemOpen
  \bibfield  {author} {\bibinfo {author} {\bibfnamefont {K.~S.}\ \bibnamefont
  {Silmore}}\ and\ \bibinfo {author} {\bibfnamefont {J.~W.}\ \bibnamefont
  {Swan}},\ }\bibfield  {title} {\bibinfo {title} {Collective mode {{Brownian}}
  dynamics: {{A}} method for fast relaxation of statistical ensembles},\ }\href
  {https://doi.org/10.1063/1.5129648} {\bibfield  {journal} {\bibinfo
  {journal} {J. Chem. Phys.}\ }\textbf {\bibinfo {volume} {152}},\ \bibinfo
  {pages} {094104} (\bibinfo {year} {2020})}\BibitemShut {NoStop}%
\bibitem [{\citenamefont {Rotne}\ and\ \citenamefont
  {Prager}(1969)}]{rotne1969}%
  \BibitemOpen
  \bibfield  {author} {\bibinfo {author} {\bibfnamefont {J.}~\bibnamefont
  {Rotne}}\ and\ \bibinfo {author} {\bibfnamefont {S.}~\bibnamefont {Prager}},\
  }\bibfield  {title} {\bibinfo {title} {Variational {{Treatment}} of
  {{Hydrodynamic Interaction}} in {{Polymers}}},\ }\href
  {https://doi.org/10.1063/1.1670977} {\bibfield  {journal} {\bibinfo
  {journal} {J. Chem. Phys.}\ }\textbf {\bibinfo {volume} {50}},\ \bibinfo
  {pages} {4831} (\bibinfo {year} {1969})}\BibitemShut {NoStop}%
\bibitem [{\citenamefont {Yamakawa}(1970)}]{yamakawa1970}%
  \BibitemOpen
  \bibfield  {author} {\bibinfo {author} {\bibfnamefont {H.}~\bibnamefont
  {Yamakawa}},\ }\bibfield  {title} {\bibinfo {title} {Transport {{Properties}}
  of {{Polymer Chains}} in {{Dilute Solution}}: {{Hydrodynamic Interaction}}},\
  }\href {https://doi.org/10.1063/1.1673799} {\bibfield  {journal} {\bibinfo
  {journal} {J. Chem. Phys.}\ }\textbf {\bibinfo {volume} {53}},\ \bibinfo
  {pages} {436} (\bibinfo {year} {1970})}\BibitemShut {NoStop}%
\bibitem [{\citenamefont {Fiore}\ \emph {et~al.}(2017)\citenamefont {Fiore},
  \citenamefont {Balboa~Usabiaga}, \citenamefont {Donev},\ and\ \citenamefont
  {Swan}}]{fiore2017}%
  \BibitemOpen
  \bibfield  {author} {\bibinfo {author} {\bibfnamefont {A.~M.}\ \bibnamefont
  {Fiore}}, \bibinfo {author} {\bibfnamefont {F.}~\bibnamefont
  {Balboa~Usabiaga}}, \bibinfo {author} {\bibfnamefont {A.}~\bibnamefont
  {Donev}},\ and\ \bibinfo {author} {\bibfnamefont {J.~W.}\ \bibnamefont
  {Swan}},\ }\bibfield  {title} {\bibinfo {title} {Rapid sampling of stochastic
  displacements in {{Brownian}} dynamics simulations},\ }\href
  {https://doi.org/10.1063/1.4978242} {\bibfield  {journal} {\bibinfo
  {journal} {J. Chem. Phys.}\ }\textbf {\bibinfo {volume} {146}},\ \bibinfo
  {pages} {124116} (\bibinfo {year} {2017})}\BibitemShut {NoStop}%
\bibitem [{\citenamefont {Chow}\ and\ \citenamefont {Saad}(2014)}]{chow2014}%
  \BibitemOpen
  \bibfield  {author} {\bibinfo {author} {\bibfnamefont {E.}~\bibnamefont
  {Chow}}\ and\ \bibinfo {author} {\bibfnamefont {Y.}~\bibnamefont {Saad}},\
  }\bibfield  {title} {\bibinfo {title} {Preconditioned {{Krylov Subspace
  Methods}} for {{Sampling Multivariate Gaussian Distributions}}},\ }\href
  {https://doi.org/10.1137/130920587} {\bibfield  {journal} {\bibinfo
  {journal} {SIAM J. Sci. Comput.}\ }\textbf {\bibinfo {volume} {36}},\
  \bibinfo {pages} {A588} (\bibinfo {year} {2014})}\BibitemShut {NoStop}%
\bibitem [{\citenamefont {Swan}\ and\ \citenamefont {Wang}(2016)}]{swan2016}%
  \BibitemOpen
  \bibfield  {author} {\bibinfo {author} {\bibfnamefont {J.~W.}\ \bibnamefont
  {Swan}}\ and\ \bibinfo {author} {\bibfnamefont {G.}~\bibnamefont {Wang}},\
  }\bibfield  {title} {\bibinfo {title} {Rapid calculation of hydrodynamic and
  transport properties in concentrated solutions of colloidal particles and
  macromolecules},\ }\href {https://doi.org/10.1063/1.4939581} {\bibfield
  {journal} {\bibinfo  {journal} {Physics of Fluids}\ }\textbf {\bibinfo
  {volume} {28}},\ \bibinfo {pages} {011902} (\bibinfo {year}
  {2016})}\BibitemShut {NoStop}%
\bibitem [{\citenamefont {Kamal}\ \emph {et~al.}(2021)\citenamefont {Kamal},
  \citenamefont {Gravelle},\ and\ \citenamefont {Botto}}]{kamal2021}%
  \BibitemOpen
  \bibfield  {author} {\bibinfo {author} {\bibfnamefont {C.}~\bibnamefont
  {Kamal}}, \bibinfo {author} {\bibfnamefont {S.}~\bibnamefont {Gravelle}},\
  and\ \bibinfo {author} {\bibfnamefont {L.}~\bibnamefont {Botto}},\ }\bibfield
   {title} {\bibinfo {title} {Effect of hydrodynamic slip on the rotational
  dynamics of a thin {{Brownian}} platelet in shear flow},\ }\bibfield
  {journal} {\bibinfo  {journal} {Journal of Fluid Mechanics}\ }\textbf
  {\bibinfo {volume} {919}},\ \href {https://doi.org/10.1017/jfm.2021.327}
  {10.1017/jfm.2021.327} (\bibinfo {year} {2021})\BibitemShut {NoStop}%
\bibitem [{\citenamefont {Anderson}\ \emph {et~al.}(2020)\citenamefont
  {Anderson}, \citenamefont {Glaser},\ and\ \citenamefont
  {Glotzer}}]{anderson2020}%
  \BibitemOpen
  \bibfield  {author} {\bibinfo {author} {\bibfnamefont {J.~A.}\ \bibnamefont
  {Anderson}}, \bibinfo {author} {\bibfnamefont {J.}~\bibnamefont {Glaser}},\
  and\ \bibinfo {author} {\bibfnamefont {S.~C.}\ \bibnamefont {Glotzer}},\
  }\bibfield  {title} {\bibinfo {title} {{{HOOMD}}-blue: {{A Python}} package
  for high-performance molecular dynamics and hard particle {{Monte Carlo}}
  simulations},\ }\href {https://doi.org/10.1016/j.commatsci.2019.109363}
  {\bibfield  {journal} {\bibinfo  {journal} {Computational Materials Science}\
  }\textbf {\bibinfo {volume} {173}},\ \bibinfo {pages} {109363} (\bibinfo
  {year} {2020})}\BibitemShut {NoStop}%
\bibitem [{\citenamefont {Bishop}(2006)}]{bishop2006}%
  \BibitemOpen
  \bibfield  {author} {\bibinfo {author} {\bibfnamefont {C.~M.}\ \bibnamefont
  {Bishop}},\ }\href@noop {} {\emph {\bibinfo {title} {Pattern Recognition and
  Machine Learning}}},\ Information Science and Statistics\ (\bibinfo
  {publisher} {{Springer}},\ \bibinfo {address} {{New York}},\ \bibinfo {year}
  {2006})\BibitemShut {NoStop}%
\bibitem [{\citenamefont {Manikantan}\ and\ \citenamefont
  {Saintillan}(2015)}]{manikantan2015}%
  \BibitemOpen
  \bibfield  {author} {\bibinfo {author} {\bibfnamefont {H.}~\bibnamefont
  {Manikantan}}\ and\ \bibinfo {author} {\bibfnamefont {D.}~\bibnamefont
  {Saintillan}},\ }\bibfield  {title} {\bibinfo {title} {Buckling transition of
  a semiflexible filament in extensional flow},\ }\href
  {https://doi.org/10.1103/PhysRevE.92.041002} {\bibfield  {journal} {\bibinfo
  {journal} {Phys. Rev. E}\ }\textbf {\bibinfo {volume} {92}},\ \bibinfo
  {pages} {041002} (\bibinfo {year} {2015})}\BibitemShut {NoStop}%
\bibitem [{\citenamefont {Schroeder}\ \emph {et~al.}(2005)\citenamefont
  {Schroeder}, \citenamefont {Teixeira}, \citenamefont {Shaqfeh},\ and\
  \citenamefont {Chu}}]{schroeder2005}%
  \BibitemOpen
  \bibfield  {author} {\bibinfo {author} {\bibfnamefont {C.~M.}\ \bibnamefont
  {Schroeder}}, \bibinfo {author} {\bibfnamefont {R.~E.}\ \bibnamefont
  {Teixeira}}, \bibinfo {author} {\bibfnamefont {E.~S.~G.}\ \bibnamefont
  {Shaqfeh}},\ and\ \bibinfo {author} {\bibfnamefont {S.}~\bibnamefont {Chu}},\
  }\bibfield  {title} {\bibinfo {title} {Characteristic {{Periodic Motion}} of
  {{Polymers}} in {{Shear Flow}}},\ }\href
  {https://doi.org/10.1103/PhysRevLett.95.018301} {\bibfield  {journal}
  {\bibinfo  {journal} {Phys. Rev. Lett.}\ }\textbf {\bibinfo {volume} {95}},\
  \bibinfo {pages} {018301} (\bibinfo {year} {2005})}\BibitemShut {NoStop}%
\bibitem [{\citenamefont {Smith}\ \emph {et~al.}(1999)\citenamefont {Smith},
  \citenamefont {Babcock},\ and\ \citenamefont {Chu}}]{smith1999}%
  \BibitemOpen
  \bibfield  {author} {\bibinfo {author} {\bibfnamefont {D.~E.}\ \bibnamefont
  {Smith}}, \bibinfo {author} {\bibfnamefont {H.~P.}\ \bibnamefont {Babcock}},\
  and\ \bibinfo {author} {\bibfnamefont {S.}~\bibnamefont {Chu}},\ }\bibfield
  {title} {\bibinfo {title} {Single-{{Polymer Dynamics}} in {{Steady Shear
  Flow}}},\ }\href {https://doi.org/10.1126/science.283.5408.1724} {\bibfield
  {journal} {\bibinfo  {journal} {Science}\ }\textbf {\bibinfo {volume}
  {283}},\ \bibinfo {pages} {1724} (\bibinfo {year} {1999})}\BibitemShut
  {NoStop}%
\bibitem [{\citenamefont {Gerashchenko}\ and\ \citenamefont
  {Steinberg}(2006)}]{gerashchenko2006}%
  \BibitemOpen
  \bibfield  {author} {\bibinfo {author} {\bibfnamefont {S.}~\bibnamefont
  {Gerashchenko}}\ and\ \bibinfo {author} {\bibfnamefont {V.}~\bibnamefont
  {Steinberg}},\ }\bibfield  {title} {\bibinfo {title} {Statistics of
  {{Tumbling}} of a {{Single Polymer Molecule}} in {{Shear Flow}}},\ }\href
  {https://doi.org/10.1103/PhysRevLett.96.038304} {\bibfield  {journal}
  {\bibinfo  {journal} {Phys. Rev. Lett.}\ }\textbf {\bibinfo {volume} {96}},\
  \bibinfo {pages} {038304} (\bibinfo {year} {2006})}\BibitemShut {NoStop}%
\bibitem [{\citenamefont {Winkler}(2006)}]{winkler2006}%
  \BibitemOpen
  \bibfield  {author} {\bibinfo {author} {\bibfnamefont {R.~G.}\ \bibnamefont
  {Winkler}},\ }\bibfield  {title} {\bibinfo {title} {Semiflexible {{Polymers}}
  in {{Shear Flow}}},\ }\href {https://doi.org/10.1103/PhysRevLett.97.128301}
  {\bibfield  {journal} {\bibinfo  {journal} {Phys. Rev. Lett.}\ }\textbf
  {\bibinfo {volume} {97}},\ \bibinfo {pages} {128301} (\bibinfo {year}
  {2006})}\BibitemShut {NoStop}%
\bibitem [{\citenamefont {Huang}\ \emph {et~al.}(2011)\citenamefont {Huang},
  \citenamefont {Sutmann}, \citenamefont {Gompper},\ and\ \citenamefont
  {Winkler}}]{huang2011}%
  \BibitemOpen
  \bibfield  {author} {\bibinfo {author} {\bibfnamefont {C.-C.}\ \bibnamefont
  {Huang}}, \bibinfo {author} {\bibfnamefont {G.}~\bibnamefont {Sutmann}},
  \bibinfo {author} {\bibfnamefont {G.}~\bibnamefont {Gompper}},\ and\ \bibinfo
  {author} {\bibfnamefont {R.~G.}\ \bibnamefont {Winkler}},\ }\bibfield
  {title} {\bibinfo {title} {Tumbling of polymers in semidilute solution under
  shear flow},\ }\href {https://doi.org/10.1209/0295-5075/93/54004} {\bibfield
  {journal} {\bibinfo  {journal} {Europhys. Lett.}\ }\textbf {\bibinfo {volume}
  {93}},\ \bibinfo {pages} {54004} (\bibinfo {year} {2011})}\BibitemShut
  {NoStop}%
\bibitem [{\citenamefont {Dalal}\ \emph {et~al.}(2012)\citenamefont {Dalal},
  \citenamefont {Hoda},\ and\ \citenamefont {Larson}}]{dalal2012}%
  \BibitemOpen
  \bibfield  {author} {\bibinfo {author} {\bibfnamefont {I.~S.}\ \bibnamefont
  {Dalal}}, \bibinfo {author} {\bibfnamefont {N.}~\bibnamefont {Hoda}},\ and\
  \bibinfo {author} {\bibfnamefont {R.~G.}\ \bibnamefont {Larson}},\ }\bibfield
   {title} {\bibinfo {title} {Multiple regimes of deformation in shearing flow
  of isolated polymers},\ }\href {https://doi.org/10.1122/1.3679461} {\bibfield
   {journal} {\bibinfo  {journal} {Journal of Rheology}\ }\textbf {\bibinfo
  {volume} {56}},\ \bibinfo {pages} {305} (\bibinfo {year} {2012})}\BibitemShut
  {NoStop}%
\bibitem [{\citenamefont {Harasim}\ \emph {et~al.}(2013)\citenamefont
  {Harasim}, \citenamefont {Wunderlich}, \citenamefont {Peleg}, \citenamefont
  {Kr{\"o}ger},\ and\ \citenamefont {Bausch}}]{harasim2013}%
  \BibitemOpen
  \bibfield  {author} {\bibinfo {author} {\bibfnamefont {M.}~\bibnamefont
  {Harasim}}, \bibinfo {author} {\bibfnamefont {B.}~\bibnamefont {Wunderlich}},
  \bibinfo {author} {\bibfnamefont {O.}~\bibnamefont {Peleg}}, \bibinfo
  {author} {\bibfnamefont {M.}~\bibnamefont {Kr{\"o}ger}},\ and\ \bibinfo
  {author} {\bibfnamefont {A.~R.}\ \bibnamefont {Bausch}},\ }\bibfield  {title}
  {\bibinfo {title} {Direct {{Observation}} of the {{Dynamics}} of
  {{Semiflexible Polymers}} in {{Shear Flow}}},\ }\href
  {https://doi.org/10.1103/PhysRevLett.110.108302} {\bibfield  {journal}
  {\bibinfo  {journal} {Phys. Rev. Lett.}\ }\textbf {\bibinfo {volume} {110}},\
  \bibinfo {pages} {108302} (\bibinfo {year} {2013})}\BibitemShut {NoStop}%
\bibitem [{\citenamefont {Chertkov}\ \emph {et~al.}(2005)\citenamefont
  {Chertkov}, \citenamefont {Kolokolov}, \citenamefont {Lebedev},\ and\
  \citenamefont {Turitsyn}}]{chertkov2005}%
  \BibitemOpen
  \bibfield  {author} {\bibinfo {author} {\bibfnamefont {M.}~\bibnamefont
  {Chertkov}}, \bibinfo {author} {\bibfnamefont {I.}~\bibnamefont {Kolokolov}},
  \bibinfo {author} {\bibfnamefont {V.}~\bibnamefont {Lebedev}},\ and\ \bibinfo
  {author} {\bibfnamefont {K.}~\bibnamefont {Turitsyn}},\ }\bibfield  {title}
  {\bibinfo {title} {Polymer statistics in a random flow with mean shear},\
  }\href {https://doi.org/10.1017/S0022112005003939} {\bibfield  {journal}
  {\bibinfo  {journal} {J. Fluid Mech.}\ }\textbf {\bibinfo {volume} {531}},\
  \bibinfo {pages} {251} (\bibinfo {year} {2005})}\BibitemShut {NoStop}%
\bibitem [{\citenamefont {Kumar}\ and\ \citenamefont
  {Yildirim}(2005)}]{kumar2005}%
  \BibitemOpen
  \bibfield  {author} {\bibinfo {author} {\bibfnamefont {P.}~\bibnamefont
  {Kumar}}\ and\ \bibinfo {author} {\bibfnamefont {E.~A.}\ \bibnamefont
  {Yildirim}},\ }\bibfield  {title} {\bibinfo {title} {Minimum-{{Volume
  Enclosing Ellipsoids}} and {{Core Sets}}},\ }\href
  {https://doi.org/10.1007/s10957-005-2653-6} {\bibfield  {journal} {\bibinfo
  {journal} {J. Optim. Theory Appl.}\ }\textbf {\bibinfo {volume} {126}},\
  \bibinfo {pages} {1} (\bibinfo {year} {2005})}\BibitemShut {NoStop}%
\bibitem [{\citenamefont {Pennec}(2006)}]{pennec2006}%
  \BibitemOpen
  \bibfield  {author} {\bibinfo {author} {\bibfnamefont {X.}~\bibnamefont
  {Pennec}},\ }\bibfield  {title} {\bibinfo {title} {Intrinsic {{Statistics}}
  on {{Riemannian Manifolds}}: {{Basic Tools}} for {{Geometric
  Measurements}}},\ }\href {https://doi.org/10.1007/s10851-006-6228-4}
  {\bibfield  {journal} {\bibinfo  {journal} {J. Math. Imaging Vis.}\ }\textbf
  {\bibinfo {volume} {25}},\ \bibinfo {pages} {127} (\bibinfo {year}
  {2006})}\BibitemShut {NoStop}%
\bibitem [{\citenamefont {Kim}\ and\ \citenamefont {Karrila}(2005)}]{kim2005}%
  \BibitemOpen
  \bibfield  {author} {\bibinfo {author} {\bibfnamefont {S.}~\bibnamefont
  {Kim}}\ and\ \bibinfo {author} {\bibfnamefont {S.~J.}\ \bibnamefont
  {Karrila}},\ }\href@noop {} {\emph {\bibinfo {title} {Microhydrodynamics:
  Principles and Selected Applications}}}\ (\bibinfo  {publisher} {{Dover
  Publications}},\ \bibinfo {address} {{Mineola, N.Y}},\ \bibinfo {year}
  {2005})\BibitemShut {NoStop}%
\bibitem [{\citenamefont {Guazzelli}\ and\ \citenamefont
  {Morris}(2011)}]{guazzelli2011}%
  \BibitemOpen
  \bibfield  {author} {\bibinfo {author} {\bibfnamefont {{\'E}.}~\bibnamefont
  {Guazzelli}}\ and\ \bibinfo {author} {\bibfnamefont {J.~F.}\ \bibnamefont
  {Morris}},\ }\href@noop {} {\emph {\bibinfo {title} {{A Physical Introduction
  to Suspension Dynamics}}}}\ (\bibinfo  {publisher} {{Cambridge University
  Press}},\ \bibinfo {year} {2011})\BibitemShut {NoStop}%
\bibitem [{\citenamefont {Kirkwood}\ and\ \citenamefont
  {Riseman}(1948)}]{kirkwood1948}%
  \BibitemOpen
  \bibfield  {author} {\bibinfo {author} {\bibfnamefont {J.~G.}\ \bibnamefont
  {Kirkwood}}\ and\ \bibinfo {author} {\bibfnamefont {J.}~\bibnamefont
  {Riseman}},\ }\bibfield  {title} {\bibinfo {title} {The {{Intrinsic
  Viscosities}} and {{Diffusion Constants}} of {{Flexible Macromolecules}} in
  {{Solution}}},\ }\href {https://doi.org/10.1063/1.1746947} {\bibfield
  {journal} {\bibinfo  {journal} {The Journal of Chemical Physics}\ }\textbf
  {\bibinfo {volume} {16}},\ \bibinfo {pages} {565} (\bibinfo {year}
  {1948})}\BibitemShut {NoStop}%
\bibitem [{\citenamefont {Zwanzig}\ \emph {et~al.}(1968)\citenamefont
  {Zwanzig}, \citenamefont {Kiefer},\ and\ \citenamefont
  {Weiss}}]{zwanzig1968}%
  \BibitemOpen
  \bibfield  {author} {\bibinfo {author} {\bibfnamefont {R.}~\bibnamefont
  {Zwanzig}}, \bibinfo {author} {\bibfnamefont {J.}~\bibnamefont {Kiefer}},\
  and\ \bibinfo {author} {\bibfnamefont {G.~H.}\ \bibnamefont {Weiss}},\
  }\bibfield  {title} {\bibinfo {title} {On the {{Validity}} of the
  {{Kirkwood}}-{{Riseman Theory}}},\ }\href
  {https://doi.org/10.1073/pnas.60.2.381} {\bibfield  {journal} {\bibinfo
  {journal} {PNAS}\ }\textbf {\bibinfo {volume} {60}},\ \bibinfo {pages} {381}
  (\bibinfo {year} {1968})}\BibitemShut {NoStop}%
\bibitem [{\citenamefont {Nir}\ \emph {et~al.}(1975)\citenamefont {Nir},
  \citenamefont {Weinberger},\ and\ \citenamefont {Acrivos}}]{nir1975}%
  \BibitemOpen
  \bibfield  {author} {\bibinfo {author} {\bibfnamefont {A.}~\bibnamefont
  {Nir}}, \bibinfo {author} {\bibfnamefont {H.~F.}\ \bibnamefont
  {Weinberger}},\ and\ \bibinfo {author} {\bibfnamefont {A.}~\bibnamefont
  {Acrivos}},\ }\bibfield  {title} {\bibinfo {title} {Variational inequalities
  for a body in a viscous shearing flow},\ }\href
  {https://doi.org/10.1017/S0022112075001206} {\bibfield  {journal} {\bibinfo
  {journal} {Journal of Fluid Mechanics}\ }\textbf {\bibinfo {volume} {68}},\
  \bibinfo {pages} {739} (\bibinfo {year} {1975})}\BibitemShut {NoStop}%
\bibitem [{\citenamefont {Poulin}\ \emph {et~al.}(2016)\citenamefont {Poulin},
  \citenamefont {Jalili}, \citenamefont {Neri}, \citenamefont {Nallet},
  \citenamefont {Divoux}, \citenamefont {Colin}, \citenamefont {Aboutalebi},
  \citenamefont {Wallace},\ and\ \citenamefont {Zakri}}]{poulin2016}%
  \BibitemOpen
  \bibfield  {author} {\bibinfo {author} {\bibfnamefont {P.}~\bibnamefont
  {Poulin}}, \bibinfo {author} {\bibfnamefont {R.}~\bibnamefont {Jalili}},
  \bibinfo {author} {\bibfnamefont {W.}~\bibnamefont {Neri}}, \bibinfo {author}
  {\bibfnamefont {F.}~\bibnamefont {Nallet}}, \bibinfo {author} {\bibfnamefont
  {T.}~\bibnamefont {Divoux}}, \bibinfo {author} {\bibfnamefont
  {A.}~\bibnamefont {Colin}}, \bibinfo {author} {\bibfnamefont {S.~H.}\
  \bibnamefont {Aboutalebi}}, \bibinfo {author} {\bibfnamefont
  {G.}~\bibnamefont {Wallace}},\ and\ \bibinfo {author} {\bibfnamefont
  {C.}~\bibnamefont {Zakri}},\ }\bibfield  {title} {\bibinfo {title}
  {Superflexibility of graphene oxide},\ }\href
  {https://doi.org/10.1073/pnas.1605121113} {\bibfield  {journal} {\bibinfo
  {journal} {Proc. Natl. Acad. Sci. U.S.A.}\ }\textbf {\bibinfo {volume}
  {113}},\ \bibinfo {pages} {11088} (\bibinfo {year} {2016})}\BibitemShut
  {NoStop}%
\bibitem [{\citenamefont {Zhang}\ \emph {et~al.}(2018)\citenamefont {Zhang},
  \citenamefont {Tang}, \citenamefont {Wang}, \citenamefont {Zhang},
  \citenamefont {Guo},\ and\ \citenamefont {Zhang}}]{zhang2018}%
  \BibitemOpen
  \bibfield  {author} {\bibinfo {author} {\bibfnamefont {X.}~\bibnamefont
  {Zhang}}, \bibinfo {author} {\bibfnamefont {Y.}~\bibnamefont {Tang}},
  \bibinfo {author} {\bibfnamefont {X.}~\bibnamefont {Wang}}, \bibinfo {author}
  {\bibfnamefont {J.}~\bibnamefont {Zhang}}, \bibinfo {author} {\bibfnamefont
  {D.}~\bibnamefont {Guo}},\ and\ \bibinfo {author} {\bibfnamefont
  {X.}~\bibnamefont {Zhang}},\ }\bibfield  {title} {\bibinfo {title}
  {Dispersion and {{Rheological Properties}} of {{Aqueous Graphene
  Suspensions}} in {{Presence}} of {{Nanocrystalline Cellulose}}},\ }\href
  {https://doi.org/10.1007/s10924-018-1237-0} {\bibfield  {journal} {\bibinfo
  {journal} {J Polym Environ}\ }\textbf {\bibinfo {volume} {26}},\ \bibinfo
  {pages} {3502} (\bibinfo {year} {2018})}\BibitemShut {NoStop}%
\end{thebibliography}%

\end{document}